\documentclass[12pt, a4paper]{article}
\usepackage[utf8]{inputenc}
\usepackage{dcolumn,lscape}
\usepackage{graphicx}
\usepackage{natbib} 
\usepackage{url} 
\usepackage[table]{xcolor} 
\usepackage{amsmath,amssymb}
\usepackage{hhline}
\usepackage{graphicx}
\usepackage{arydshln}
\usepackage[format=hang,labelfont=bf,textfont=small,singlelinecheck=false,justification=raggedright,margin={12pt,12pt},figurename=Fig.]{caption}
\usepackage{titlesec}
\usepackage{textcomp}
\usepackage{amsmath,amssymb,dsfont, mathtools}
\usepackage{makecell}
\usepackage{multirow}
\usepackage{exscale,longtable,multicol,dcolumn,tabularx,bm}
\usepackage{adjustbox}
\usepackage{neuralnetwork}
\usepackage{amsmath,longtable,multicol,dcolumn,tabularx,graphicx,amssymb}
\usepackage{exscale,amsthm,multirow,rotating}
\usepackage{bbm}
\usepackage{array}
\usepackage{comment}
\usepackage{tikz,dsfont,float}
\usepackage[utf8]{inputenc}
\usepackage{hyperref}
\hypersetup{
    colorlinks=true,
    citecolor = black,
    linkcolor=blue,
    filecolor=magenta,      
    urlcolor= blue 
}
\usepackage{bbm}
\usepackage{tikz,dsfont,float}
\usepackage{mathptmx} 
\usepackage{mathrsfs}
\usepackage{booktabs} 
\usepackage{subcaption}

\usepackage{lscape}

\usetikzlibrary{calc}
\usetikzlibrary{positioning}

\usepackage{graphicx}
\usepackage{amsmath,amssymb,dsfont, mathtools}
\usepackage{natbib}

\usepackage{makecell}
\usepackage{multirow}
\usepackage{exscale,longtable,multicol,dcolumn,tabularx,bm}
\usepackage{adjustbox}
\usepackage{booktabs}
\usepackage{mathptmx}      
\usepackage{mathrsfs}

\usepackage{xcolor, ulem}

\usepackage{algorithm}
\makeatletter
  \renewcommand{\ALG@name}{Procedure}
 
  \makeatother

\usepackage{enumitem}
\setlist[enumerate]{left=0.0em}

\begin{document}

\def\spacingset#1{\renewcommand{\baselinestretch}%
{#1}\small\normalsize} \spacingset{1}

  \title{\bf Statistical monitoring of European cross-border physical electricity flows using novel temporal edge network processes}
  \author{Anna Malinovskaya\\
  	\small{Leibniz University Hannover, Germany}\\
  	Rebecca Killick\\
  	\small{Lancaster University, United Kingdom}\\
  	Kathryn Leeming\\
  	\small{Leicestershire, United Kingdom}\\
  	Philipp Otto\\
  	\small{University of Glasgow, United Kingdom}}
  \maketitle
\begin{abstract}
Conventional modelling of networks evolving in time focuses on capturing variations in the network structure. However, the network might be static from the origin or experience only deterministic, regulated changes in its structure, providing either a physical infrastructure or a specified connection arrangement for some other processes. Thus, to detect change in its exploitation, we need to focus on the processes happening on the network. In this work, we present the concept of monitoring random Temporal Edge Network (TEN) processes that take place on the edges of a graph having a fixed structure. Our framework is based on the Generalized Network Autoregressive statistical models with time-dependent exogenous variables (GNARX models) and Cumulative Sum (CUSUM) control charts.  To demonstrate its effective detection of various types of change, we conduct a simulation study and monitor the real-world data of cross-border physical electricity flows in Europe.

\end{abstract}

\noindent%
{\it Keywords:} CUSUM Control Charts; GNARX Model; Network Modelling; Network Monitoring.
\vfill

 \newpage
\spacingset{1.45}

\section{Introduction}\label{sec:intro}
Network analysis is an important statistical inference tool in various disciplines including economics, social sciences, and chemistry \citep{de2007debt, khrabrov2010discovering, mcdermott2021graph}; It provides insight into the network structure and beyond. Usually, the underlying network structure is considered to be random, meaning that both nodes and edges may appear and disappear over time. However, processes also originate in network structures with limited or no temporal dynamics. In both cases, to understand whether a change has occurred we need to inspect the data over time. This procedure is also known as \textit{network monitoring}. Hence, to distinguish the case when the network is considered to be a random observation from the case when the network structure is deterministic but the process on it is random, we utilise the terminology \textit{random network monitoring} and \textit{fixed network monitoring} introduced in \cite{stevens2021foundations}.

Usually, the modelling and monitoring of networks are focused on detecting changes and anomalies reflected in the geometric properties of a graph, falling under the type of \textit{random network monitoring}. Illustrative studies include the monitoring of e-mail communication and the daily flights within a country for detecting unusual communication patterns (cf. \citealp{perry2020ewma, malinovskaya2021online}). However, for some networks, the development of connections or inclusion of new nodes is either no longer possible or would not be of considerable monitoring interest.

In general, the integration of nodal or edge attributes for improving the modelling of the underlying network formation mechanism and in turn the network monitoring is not a new concept \citep{azarnoush2016monitoring, shaghaghi2020pca}. However, in existing research, the attributes are considered to regulate the presence or absence of an edge. 
In other words, the contextual information available either on vertices or edges is viewed as an extra dimension to the graph helping to explain the likelihood of the observed links \citep{miller2013efficient}. In contrast, we consider the process happening on edges as the primary or only source of information about the network state.

It is beneficial to distinguish general random processes we regard on the fixed network structures from specific areas of analysis of processes on networks. The first well-established perspective is the analysis of spatial point patterns on networks (cf. \citealp{baddeley2021analysing}), where the accurate location of the object on the physical network is of concern. The second area concerns the analysis of network flows (cf. \citealp{ahuja1993network, Kolaczyk2020}), where the physical constraints of a network structure and the flow itself play a vital role. In this work, however, we could think of the analysed process as being a flow or traffic with no associated constraints unless they are imposed explicitly.

Starting with the description of TEN processes in Section \ref{sec:not}, we explain our monitoring framework in Section \ref{sec:monfram}. Afterwards, we perform a simulation study by testing the proposed methodology under different anomalous scenarios in Section \ref{sec:sim}. To illustrate the importance and the idea of how to monitor the real-world TEN processes, we monitor cross-border physical electricity flows in Europe in Section \ref{sec:ele}. In the last section, we discuss further research direction and summarise the perspectives and limitations of the proposed approach.

\section{TEN Process}
\label{sec:not}
\begin{figure}
    \centering
			\includegraphics[scale=0.3, trim= 0 0 0 0,clip]{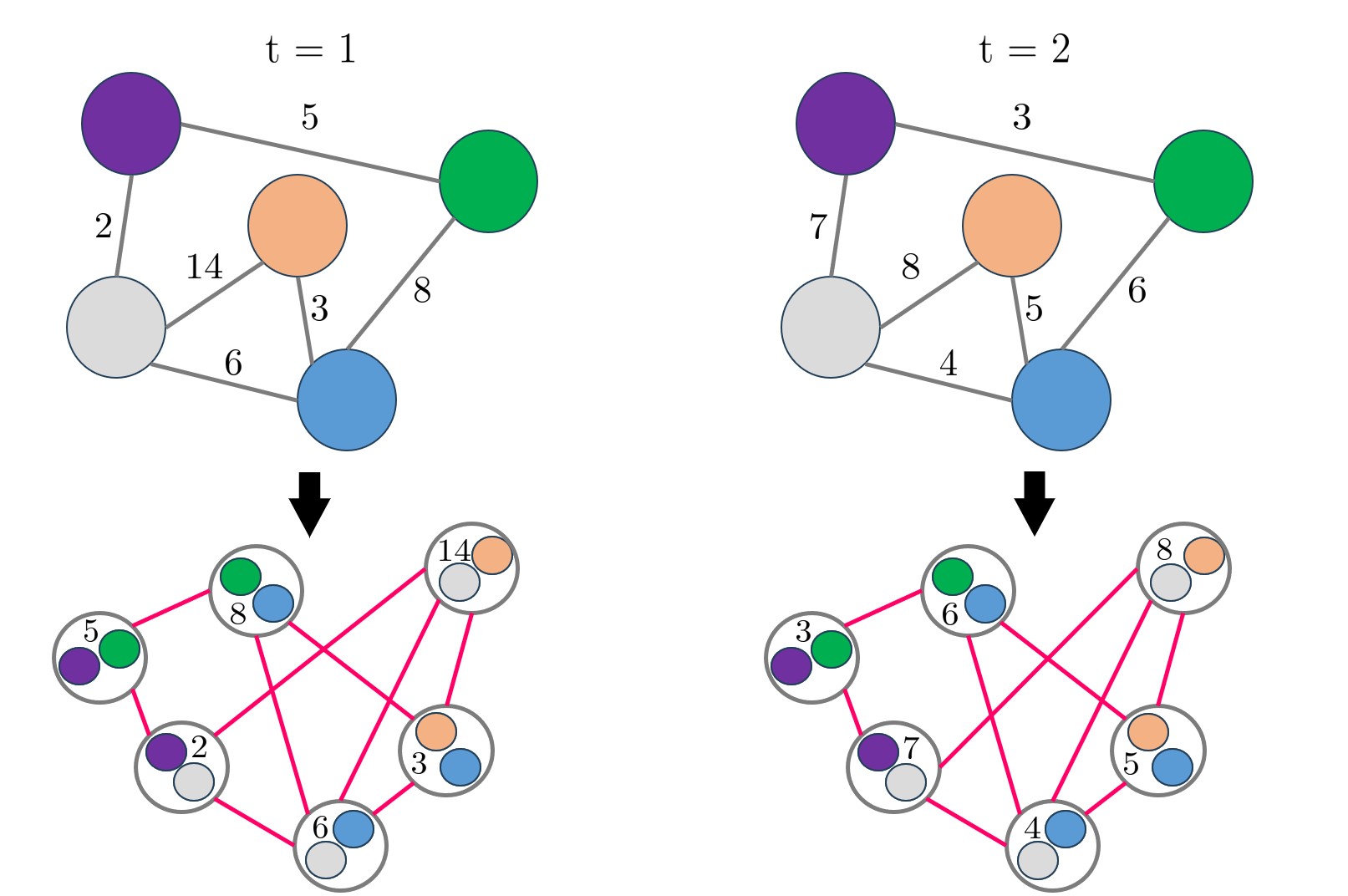}
	 \caption{\centering Visual illustration of a TEN process (top) for two time points with an adapted representation (below) described in Section \ref{subsubsec:GNARXTEN}.}
    \label{fig:TENexample}
\end{figure}

Consider a network $G = (V, E)$, where the elements of $V$ represent vertices (or nodes) and $E$ -- edges (or links). Further, we assume a fixed structure of $G$ over time described by an adjacency matrix $\bm{Y} \coloneqq (Y_{ij})_{i,j = 1, \dots, |V|}$, with $|V|$ being the number of nodes. When two vertices $i, j \in V$ are connected by an edge $e \in E$, they are called adjacent. Usually, the entries of $\bm{Y}$ are binary, i.e. $Y_{ij} = 1$ if $(i,j)\in E, \,i \ne j$, and $0$ otherwise. The graph can be directed or undirected, in the latter case is $\bm{Y}$ symmetric. To each existing edge $e \in E$ we relate time series $\{x_{e, t}\}$, where $t = 1, \dots, T$, being the attributed process of $G$. The complete representation $\bm{X} = (X_{e, t})_{e = 1, \ldots, |E|, t = 1, \ldots, T}$, where $|E|$ denotes the number of edges, together with the graph $G$ forms a dynamic Temporal Edge Network (TEN) process illustrated in Figure \ref{fig:TENexample} (top) for two time stamps. 
The following section proposes one approach to modelling and monitoring TEN processes.

\section{Network Monitoring Framework}
\label{sec:monfram}
Usually, statistical network monitoring is subdivided into two parts: 1) \textbf{Network modelling} and 2) \textbf{Process monitoring} \citep{stevens2021research}. Instead of formal network modelling, it is possible to collect network features and monitor those. For example, \cite{flossdorf2021change} comprehensively discuss which network characteristics are suitable for which types of change. In our work, prior to the monitoring procedure in Section \ref{sec:mon}, we first define a model that can be suitable for TEN processes.

\subsection{Modelling of TEN Processes}

The majority of research on network models considers the network (or graph) structure to be a random variable. Available models include Markov process models (cf. \citealp{snijders2005models}), (temporal) exponential random graph models (cf. \citealp{hanneke2010discrete}) and latent process-based models, e.g.\ (dynamic) stochastic block models (cf. \citealp{matias2017statistical}). In contrast, TEN processes assume a static structure of a network over time. One could adapt the aforementioned random network models to this context but they involve computationally intensive estimation that relies on numeric approximation. Hence, we consider alternative model forms designed for fixed network structures. 

The modelling of multivariate time series with an underlying network dependency structure is gaining popularity. \cite{JSSv096i05} present the Generalized Network Autoregressive (GNAR) framework, where a network is associated with multivariate time series and modelled as one. The GNAR models time series that occur at nodes of the network, e.g.\ growth rates of gross domestic product. The GNAR model was extended to incorporate time-dependent exogenous variables (GNARX) in \cite{nason2022quantifying}. After the introduction of the original GNARX model in Section \ref{subsubsec:GNARX}, we present its extension from nodal time series to time series on network edges in Section \ref{subsubsec:GNARXTEN}.

\subsubsection{GNARX Model}
\label{subsubsec:GNARX}
In this section, consider $\{x_{i, t}\}$ to be a time series related to each node $i$. The GNARX model $(p, \bm{s}, \bm{q})$ with $H$ exogenous regressors $\{z_{h, i, t: h = 1, \dots, H,\, i \in V,\, t = 1, \dots, T}\}$ and the autoregressive order $p$, where $(p, \bm{s}, \bm{q}) \in  \mathbb{N} \times  \mathbb{N}^p_0 \times \mathbb{N}^H_0$ holding for all vertices $i \in V$, is specified as
		\begin{equation}
			x_{i,t} = \sum_{l=1}^{p}\bigg(\alpha_{i,l}x_{i,t-l} + \sum_{r = 1}^{s_l}\beta_{l,r}\sum_{j\in \mathscr{N}^{(r)}(i)}\omega_{i,j}x_{j,t-l}\bigg) + \sum_{h=1}^{H}\sum_{q=0}^{q_h}\gamma_{h,q}z_{h,i,t-q}+\epsilon_{i, t}. 
			\label{eq:GNARX}
		\end{equation}
 The order $p$ also determines the maximum order of neighbour time lags, i.e.\ $\bm{s} = (s_1, \dots, s_p)$ with $s_l$ being the maximum stage of neighbour dependence for time lag $l$. For example, $s_1 = 2$ means that nodes depend on their first and second-stage neighbours in $G$ in the first time lag. Similarly, the maximum time lag of the $h$-th exogenous regressor $\{z_{h, i, t}\}$ is defined as $q_h$ and collectively these are $\bm{q} = (q_1, \dots, q_H)$. In case $q_h = 0$, the current value of the exogenous variable is considered at time point $t$.

 The noise is denoted by $\epsilon_{i,t}$ and is assumed to be independent and identically distributed at each vertex $i$ with mean zero and variance $\sigma^2_i$.  The parameters $\alpha_{i,l},\  \beta_{l,r},\  \gamma_{h,q} \in \mathbb{R}$ define autoregressive influence, neighbouring influence and external influence from regressors, respectively. It is possible to estimate a global-$\alpha$ model, where $\alpha_{i, l} = \alpha_l$, assuming the same autoregressive process for all nodes.  
 
\begin{figure}
    \centering
			\includegraphics[scale=0.4, trim= 1cm 1cm 1cm 1cm,clip]{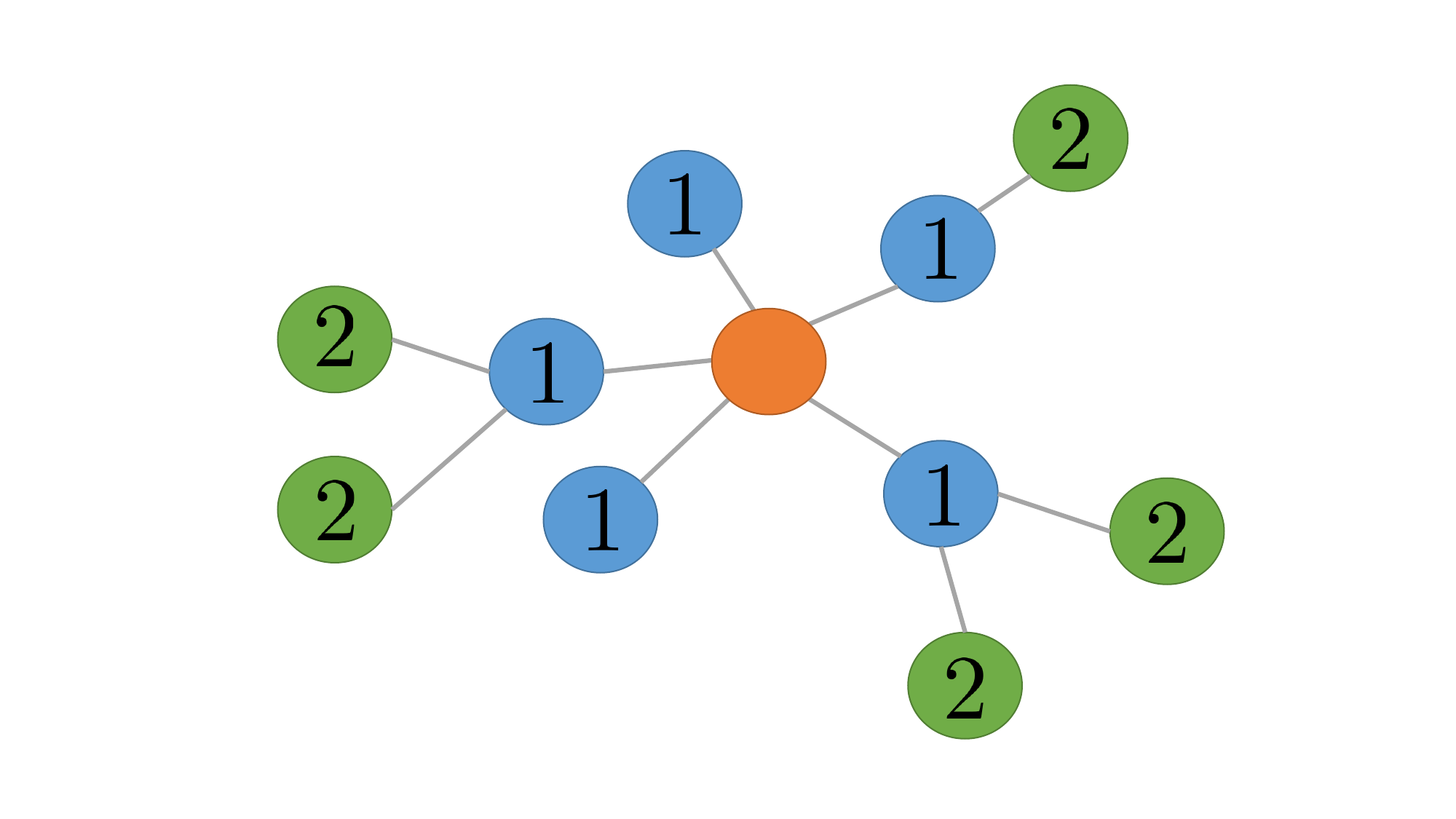}
	 \caption{\centering Stage-1 (blue) and stage-2 (green) neighbourhood of the orange node.}
    \label{fig:Neighb}
\end{figure}

 The set $\mathscr{N}^{(r)}(i)$ denotes the $r$-th stage neighbourhood set of node $i \in G$. For instance, the stage-1 neighbours of a node $i \in V$ are the adjacent nodes $j \in V$, connected by an edge. Further, the stage-2 neighbourhood set of a node $i \in V$ is the stage-1 neighbours of the adjacent nodes $j \in V$ as can be seen in Figure \ref{fig:Neighb}. Moreover, there are weights $\omega \in [0, 1]$ associated with every pair of nodes that, in our case, depend on the size of the neighbour set as explained in \cite{JSSv096i05}.

By fitting the GNARX model, we obtain the estimates of the parameters $\alpha_{i,l},\  \beta_{l,r},\  \gamma_{h,q}$ that can be used for predicting the $\hat{x}_{i, t + 1}$ value from $\{x_{i, t}\}$. The one-step-ahead forecast errors are then determined as 
\begin{equation}
u_{i, t + 1} = x_{i, t + 1} - \hat{x}_{i, t + 1}
\label{eq:res}
\end{equation}
which can be utilised in our monitoring framework. However, the GNARX model is defined for time series related to nodes. Thus, we need to extend the representation of TEN processes to enable a correct application of the GNARX model for performing the monitoring.

\subsubsection{Extension to TEN Processes}
\label{subsubsec:GNARXTEN}
There are different ways of thinking about why we require an alternative representation. First, our main focus lies in discovering changes happening between and within the streams, i.e.\ a process captured on the edges. Thus, we actually indirectly consider it to be a network not of single nodes anymore but of pairs of nodes.  Second, the influence from node to node and the influence from one stream to its neighbouring stream portray two distinct standpoints. In the case of the dependency between two traffics, we have more than two nodes involved and other underlying mechanisms that determine the exchange and its strength. Therefore, we need a different adjacency matrix to describe this novel dependency structure.

For distinguishing between $G$ and $\bm{Y}$, and the novel representation of the TEN process as $G'$, we introduce notation for the new adjacency matrix as $\bm{Y}'$. Now, as displayed in Figure \ref{fig:TENexample} (below), the edges $e \in E$ from the original graph $G$ become vertices $\iota \in V'$ with the number of nodes being $|V'| = |E|$ so that the time series $\{x_{e, t}\}$ placed on edges $e \in E$  are now placed on the nodes, becoming $\{x_{\iota, t}\}$. Consequently, we would also have new edges $\xi \in E'$. Considering the connectivity structure captured by the adjacency matrix,  without any expert knowledge, the natural choice for creating $\bm{Y}'$ would be to use a model for generating random graphs. A potential candidate is Erd\H{o}s-R\'enyi model, where all graphs on a fixed vertex set with a fixed number of edges are equally likely. Another possibility is to consider sampling from a stochastic block model which produces graphs with a community structure. For example, links within a community may be more probable than between communities. Both of those options are tested in Section \ref{sec:sim}. 

There is, however, an additional approach to choosing a new connectivity structure by referring to other types of graphs from graph theory. In this case, we use a deterministic approach to construct $\bm{Y}'$. One relevant representation technique is the line (also known as edge-to-vertex dual) graph $L(G)$ of $G$. Here, $L(G)$ is constructed on $E$, where $e \in E$ are adjacent as nodes $\iota \in V'$ if and only if they share a common node as edges in $G$ \citep{Diestel2017}. In Figure \ref{fig:TENexample} (below) this connectivity structure is adapted to the original TEN process shown above. Such formation can suit well real-world applications as it offers a more logical structure in case of limited knowledge about any existing communities or insufficiency of a connectivity structure generated from a random graph model.

After redefining the representation of TEN processes, we can proceed with their modelling by applying the GNARX model and their monitoring using a residual-based control chart introduced in the next section.

\subsection{Monitoring of TEN Processes}
\label{sec:mon}

Recall that for network monitoring we need a model and a monitoring procedure. For monitoring, it is vital to determine the framework in which we are aiming to perform it. One possibility is to monitor the estimates of the model parameters although this is computationally intensive (cf. \citealp{malinovskaya2021online}). A more efficient alternative is to perform monitoring based on residuals. For instance, \cite{miller2013efficient} compute graph residuals as the difference between the observed graph and its expected value. In modelling the TEN with the GNARX model (see Equation \ref{eq:res}), it is uncomplicated to calculate the residuals as the deviations from the current observation and the GNARX prediction. As soon as a change occurs, we would expect the residuals to increase indicating a lack of fit. According to \cite{alexopoulos2004spc}, who offers a detailed introduction to Statistical Process Monitoring (SPM) and especially to the forecast-based monitoring methods, when the model or prediction is accurate, the prediction errors are uncorrelated. Thus, we can apply traditional SPM techniques such as control charts to these forecast errors. A technique that is particularly suitable for testing whether the current residual-based statistic is anomalous is a control chart.

Belonging to the methods of SPM (cf. \citealp{montgomery2012statistical}), control charts are both straightforward to implement and interpret. \cite{grundy2021aspects} proposes a framework for monitoring forecast models, showing that changes in complex data are reflected in the forecast errors. They utilise the Cumulative Sum (CUSUM) control chart, and we extend their approach to monitoring residuals of the GNARX model for TENs as shown subsequently. A comprehensive overview of residual control charts and their comparison to other types of control charts is provided in \cite{knoth2004control} and \cite{jensen2006effects}. 

Applying control charts in an online manner, i.e. performing a sequential change point detection, at each time point $t$ we test whether the process remains in control. Hence, the specific hypothesis as well as the definition of a change point are clarified in Section \ref{sub:hypothesis} before coming to the description of the implemented control chart in Section \ref{sub:cpt}.

\subsubsection{Hypothesis Formulation}
\label{sub:hypothesis}

As we base our monitoring part on the residuals obtained by finding the difference between the actual process and the predictor at each flow (Equation \ref{eq:res}), it is vital to obtain an accurate model before monitoring starts. Our objective is then to test 
\begin{equation*}
H_{0, t}: \, \text{The observed TEN process coincides with the fitted GNARX model}
\end{equation*}
\vspace{0.2cm}
against the alternative
\vspace{-0.3cm}
\begin{equation*}
\hspace{1.3cm} H_{1, t}: \, \text{The observed TEN process does not coincide with} \text{the fitted GNARX model}
\end{equation*}
for each $t$. Now the question arises of which variable to use for the test statistic that could provide us with the information about when $H_0$ is violated. The change point $\tau$ is defined as 
\begin{equation*}\label{eq:cp_model}
\bm{x}_t \quad \sim \quad \left\{ \begin{array}{cc}
F(\bm{\mu},  \, \Sigma) &  \text{ if } t < \tau \\
F_\tau(\bm{\mu}_\tau,  \, \Sigma_\tau) & \text{ if } t \geq \tau \\
\end{array} \right. \, ,
\end{equation*}
where $\bm{x}_t = (x_{1, t}, \dots, x_{|V'|, t})'$, $\bm{\mu}$ and $\Sigma$ define the mean vector and variance-covariance matrix of the network flow distribution $F$ in our case, respectively. However, the change in the mean or/and in the variance of the raw data would lead to the respective changes in the forecast errors \citep{grundy2021aspects}. Hence, by constructing a test statistic based on $\bm{u}_t = (u_{1, t}, \dots, u_{|V'|, t})'$, we can accordingly determine the time point $\tau$ when the change has occurred. Following, we introduce the change point detection framework specified by \cite{grundy2021aspects} and discuss the respective test statistic for applying residual-based CUSUM control charts.

\subsubsection{Residual-Based CUSUM Control Chart}
\label{sub:cpt}

Before describing the technicalities of the considered control chart, it is worth beginning with an explanation of its key elements and framework for application in practice. In general, a control chart, being a graphical tool for detecting unusual variations of the process, consists of three components. These are a central line ($CL$) for plotting the average of the process, an upper control line for the upper control limit and a lower control line for the lower control limit. When the process considerably deviates from its expected state, the control statistic starts crossing the limits, indicating a substantial change beyond random variation. 

Usually, the are two phases involved in the implementation of control charts. Phase I corresponds to the exploration and calibration period, where we estimate the target process and calibrate the control chart, e.g. computing control limits. It is based on the assumption that the process during this phase is in control, i.e. stable, predictable and repeatable \citep{vining2009technical}. In Phase II the actual application of the control chart begins which could be viewed as a sequential implementation of the hypothesis test introduced in Section \ref{sub:hypothesis}. The more precise the estimation of the parameters in Phase I, the more reliable the performance of the control chart in Phase II. The desired operation of control charts during Phase II is a quick detection of the change point in a process when it experiences an out-of-control state, i.e. unusual variation of a process is present. At the same time, we strive for a long-running scheme without false alarms occurring, meaning the case when the process actually remains in control but the scheme signals a change. This translates to a balance between small control limits for fast detection and large control limits for small false alarm rates. 

In real-world applications, it is usually unknown what properties, e.g. mean or variance, of a process may change. Thus, an important criterion for selecting a suitable control chart is its ability to track many types of change simultaneously. Another point considers the decision about what exactly to monitor -- the data itself or some process related to it. As discussed by \cite{grundy2021aspects}, the monitoring of the data directly might deteriorate in case of temporal dependency, complex trends or seasonality effects within its structure. However, monitoring forecast residuals of the process omits these issues and is capable of reflecting changes in either the process mean or the process variance.

As under the real settings, it is usually unknown whether the change occurs in a mean and/or variance of the process, we utilise a more general type of Page's CUSUM detector, which is based on the (centred) squared data and is able to detect a combination of the mean and/or variance change in the original data. For monitoring forecast errors $u_{\iota,t}$ obtained for each flow $\iota$, \cite{grundy2021aspects} adapts Page's CUSUM test statistic  (cf. \citealt{page1954continuous}) to the centered squared forecast errors as follows
\begin{equation}
    Q_{\iota}(m, k) = \sum_{t = m + 1}^{m + k}(u_{\iota,t} - \hat{b})^2 - \frac{k}{m}\sum_{t=1}^{m}(u_{\iota,t} - \hat{b}),
\end{equation}
where $m$ corresponds to the length of Phase I, $k$ is the current time point in Phase II and $\hat{b}$ is the mean estimate of the forecast errors computed from Phase I.  From that, we compute the control chart statistic as 
\begin{equation}
    D_{\iota}(m, k) = \max_{0\leq a \leq k}|Q(m, k) - Q(m, a)|,
    \label{eq:chartstat}
\end{equation}
and the corresponding upper control limit is given by 
\begin{equation}
    UCL = \hat{\sigma}_{\iota}\zeta_{\alpha}g(m, k, \nu),
    \label{eq:ucl}
\end{equation}
with $\zeta_{\alpha}$ being the critical value of the distribution with the significance level $\alpha$ and $\hat{\sigma}_{\iota}$ the estimate of the standard deviation of the centred squared forecast errors belonging to the flow $\iota$. The part $g(m, k, \nu)$ defines the weight function with the tuning parameter $\nu$, both are described in detail in \cite{horvath2004monitoring} and \cite{grundy2021aspects}. In this work, we consider $\nu = 0$. The current time point $k = \tau$, i.e. the control chart detects a change point if $D_\iota(m, k) \leq UCL$. It is worth noting that no lower control limit is defined as the control statistic obtains non-negative values only and, therefore, has its natural lower control limit which is zero.

\subsubsection{Monitoring Scope and Performance Function}

The considered CUSUM chart belongs to the univariate control charts, meaning that only one process variable is monitored by the scheme. In our case, it means that we simultaneously implement $n = |V'|$ control charts to monitor the TEN process completely. Consequently, it is also possible to perform local monitoring and control the behaviour of only one essential flow from a complete TEN process. 

In this work, however, we are interested in proposing a monitoring procedure suitable for tracking the complete TEN process. Thus, as a performance metric, we introduce a cumulative intensity function of a change that presents a cumulative sum of flows that triggered a change at a time point $t$
\begin{equation}
    I_{\bm{X}} (T) = \sum_{t = 1}^{T}\frac{\sum_{\iota = 1}^{n}{\mathbbm{1}_{[UCL, \infty)}[D_{\iota}(m, k)]}}{n}.
    \label{eq:intensity}
\end{equation}
If in one flow $\iota$ a change was detected, the monitoring of this flow stops, resulting in $\max(I) = 1$. Thinking about how a signal is produced by computing $I_{\bm{X}} (T)$, we need to define a suitable threshold $W$ based on expert knowledge or system requirements.

In the subsequent section, we illustrate a simulation study that handles different cases of changes as well as compares monitoring results with respect to the selected model for constructing the adjacency matrix $\bm{Y}'$.

\section{Simulation Study}
\label{sec:sim}

\begin{figure}
\vspace{-2cm}
    \centering
			\includegraphics[scale=0.4, trim= 0 0 0 0,clip]{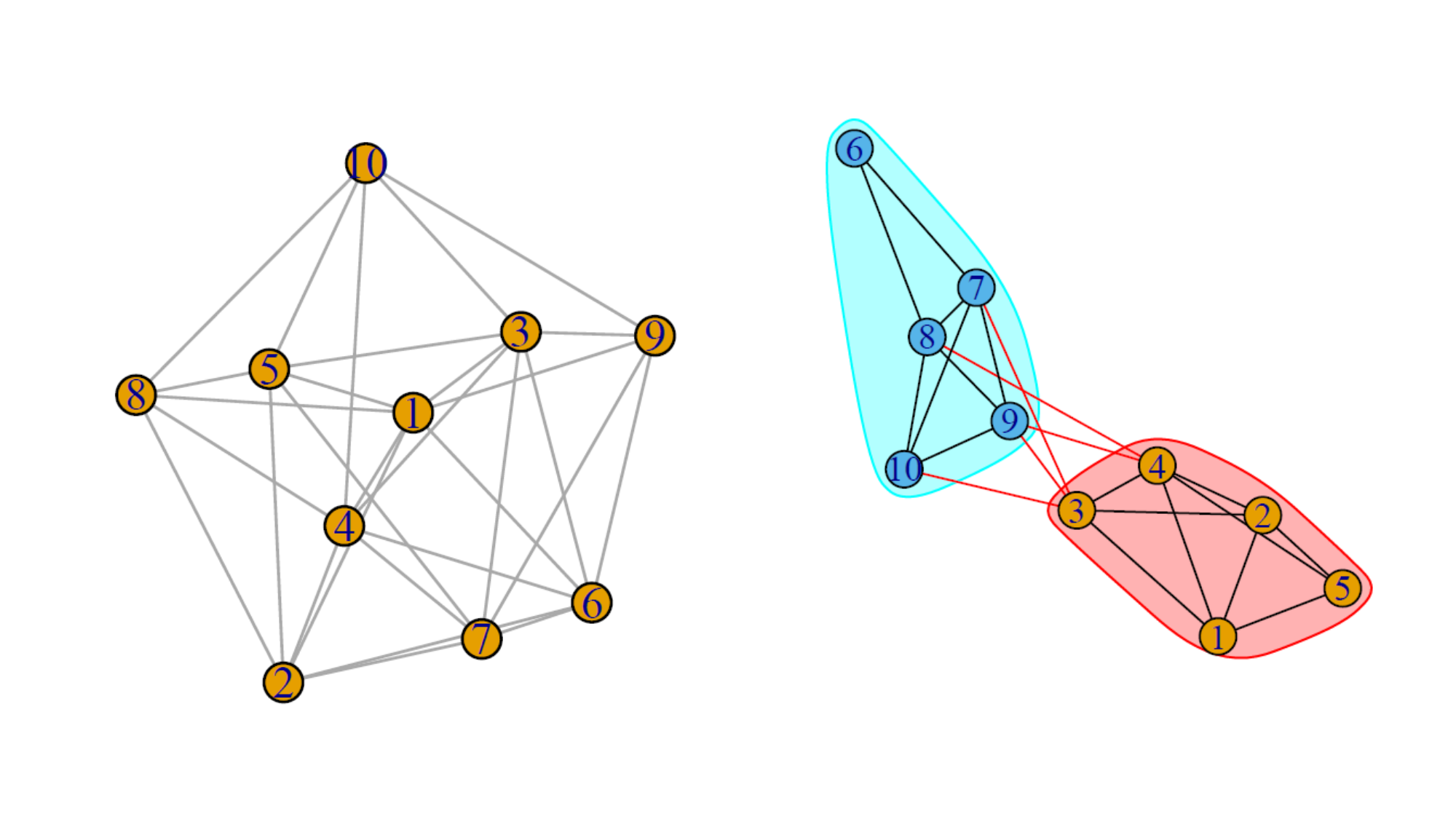}
	 \caption{\centering Representation of two different connectivity types: A graph structure generated by an Erd\H{o}s-R\'enyi model (left) vs. by a SBM (right).}
    \label{fig:Adj}
\end{figure}

The reason to conduct a simulation study is twofold: First, we aim to quantify the detection speed of changes in the effects captured by the model parameters $\alpha_{i,l}$, $\beta_{l,r}$ or $\gamma_{h, q}$. Second, as mentioned in Section \ref{sec:not}, there are different possibilities to construct the adjacency matrix $\bm{Y}'$ as introduced in Section \ref{subsubsec:GNARXTEN}. Here, we examine two distinct random graph models that cover both the availability and absence of expert knowledge about suitable sampling models. If a TEN process is well understood, the application of a Stochastic Block Model (SBM), where the flows should be first subdivided into clusters, is a good choice. Alternatively, it is possible to run a clustering algorithm. If the cluster structure is not suitable or no specific knowledge about the TEN process is available, the structure of $G'$ can be sampled by applying Erd\H{o}s-R\'enyi model. An example of both structure types is presented in Figure \ref{fig:Adj}.

In the next part, we explain the design of the conducted simulation study, presenting the results as well as comparing them with regard to the generated 
 graph structure.

\subsection{Experimental Setting}
To perform a simulation study, we have to decide on the design of a TEN process, the initial parameters and the test cases. To keep the study simple for reproducing it but comprehensive for designing different anomaly types as well as testing two distinct $\bm{Y}'$ structures, we create a TEN process with 10 flows and medium connectivity. That means, for sampling $\bm{Y}'$ from the Erd\H{o}s-R\'enyi model, we set the number of nodes to be 10 and the number of connections to be 30, considering that everything else is random. For sampling from an SBM, we separate the flows into two clusters $c_1$ and $c_2$ of equal size. Additionally, we need to define the probabilities $P$ that the flows are connected within a cluster or between the clusters. 

Regarding the settings for sampling from a GNARX model, we design a process with a global autoregressive effect of order $p = 1$. Also, the stage-1 neighbourhood is considered together with two exogenous regressors and corresponding time lags $\bm{q} = (0, 0)$. To be precise, we generate separate data for estimating the GNARX model (600 observations) before designing Phase I with 200 observations and Phase II with 100 observations which is subdivided into 50 in-control and 50 out-of-control samples. The detailed description of each simulation step is presented in Procedure \ref{alg}. In total, we conduct 500 iterations for each test case, presenting the outcomes subsequently.
\begin{algorithm}[tbp]  
\caption{}
\vspace{0.2cm}
  \begin{enumerate}
  \item \textbf{Setting:} Choose the lengths of the burn-in period (300), model fitting period (600), Phase I (200) and Phase II (100) subdivided into in-control (50) and out-of-control (50) parts. Set the data-generating parameters of the GNARX model: $\alpha_{\iota,l} = \alpha = 0.2$, $\beta_{l,r} =  \beta = 0.3$ with $p = 1$ and $\bm{s} = (1)$, $ \gamma_{h,q} = \gamma_{1} = 2$ and $\gamma_{2} = 3$ with $\bm{q} = (0,0)$. Select the change: $\beta_\tau = \beta + 0.3$. 
  \vspace{-0.3cm}
    \item \textbf{TEN Generation:}
     \begin{enumerate}
     \vspace{-0.4cm}
    \item Generate the adjacency matrix $\bm{Y}'\coloneqq (Y'_{\iota \upsilon})_{\iota,\upsilon = 1, \dots, |V'|}$ either by sampling from the  Erd\H{o}s-R\'enyi model with $|V'| = 10$ and $|E'| = 30$ or the SBM with $P_{c_1c_2} = P_{c_2c_1} = 0.2$ and $P_{c_1c_1} = P_{c_2c_2} = 0.8$.
    \item For each $t = 1, \dots, 1200$ generate the GNARX process:
    \begin{enumerate}
    \item 
    Sample error values $\epsilon_{\iota, t}$, where $\epsilon_{\iota,  t} \sim \mathcal{N}(0, 1)$.
    \item Sample covariate values $z_{1,\iota, t} \sim \mathcal{N}(0, 1)$ and $z_{2, \iota, t} \sim \mathcal{N}(0, 1)$.
    
     \item For $\iota = 1$ to $\iota = |V'|$ 
     \begin{enumerate}
         \item If $t \leq 1150$, generate the data with $\alpha, \beta, \gamma_1$ and $\gamma_2$ parameters: 
         \vspace{-0.4cm}
       \[  
       	x_{\iota, t} = \alpha x_{\iota,t - 1} + \beta\sum_{\upsilon\in \mathscr{N}^{(1)}(\iota)}\omega_{\iota,\upsilon}x_{\upsilon,t-1}+  \gamma_{1} z_{1, \iota, t-1} + \gamma_{2}z_{2, \iota, t-1} +\epsilon_{\iota, t}.
       \]
    \item If $t \geq 1151$, generate the data with  $\alpha, \beta_\tau, \gamma_1, \gamma_2$ parameters having an effect either on all flows or only on flows from $c_1$, adjusting the equation in A. 
        \end{enumerate}
    \end{enumerate}
\end{enumerate}
 \vspace{-0.5cm}
  \item \textbf{TEN Modelling:}
  \begin{enumerate}
  \vspace{-0.4cm}
      \item Fit the GNARX model with $p = r = 1$ using the first 600 generated observations after discarding the burn-in period $t = 1, \dots, 300$. 
      \item Save the estimated parameters $\Tilde{\alpha}, \Tilde{\beta}, \Tilde{\gamma}_1$ and $\Tilde{\gamma}_2$.
  \end{enumerate}
   \vspace{-0.5cm}
  \item  \textbf{TEN Monitoring:} 
  \begin{enumerate}
  \vspace{-0.4cm}
  \item Phase I: Estimate the target process by using $\Tilde{\alpha}, \Tilde{\beta}, \Tilde{\gamma}_1$ and $\Tilde{\gamma}_2$ parameters to compute forecast residuals $u_{\iota, t}$ and calibrate the CUSUM control charts described in Section \ref{sub:cpt}.
  \item Phase II: Run monitoring using the calibrated CUSUM control charts and compute the cumulative intensity function provided in Equation \ref{eq:intensity}. Save the time stamp $t$ when a change is detected. 
  \end{enumerate}
  \vspace{-0.3cm}
  \end{enumerate}
  \caption{Simulation steps for one iteration for the test case $\beta + 0.3$}
  \label{alg}
\end{algorithm}

\subsection{Results and Comparison of Connectivity Structures}
\label{subsec:simresults}

As explained in Section \ref{sec:mon}, we aim to detect changes in the whole TEN process, i.e. the anomalous network states. Therefore, we compute the cumulative change intensity given in Equation \ref{eq:intensity}. For summarising the performance in the simulation study, for each time point in Phase II we average the values of the cumulative change intensity function, obtaining a mean cumulative change intensity. It is worth noting that we do not define a threshold $W$ here, focusing on the general detection capability of the approach as well as fluctuation in the performance. It is important to note that as soon as the control chart signals for a particular flow, the monitoring of that flow stops. It means, that no multiple change points for the same flow are possible.

For each parameter $\Tilde{\alpha}, \Tilde{\beta}$ and $\Tilde{\gamma}_1$ we have three test cases of changes whose values gradually increase. Comparing the different types of sampling the adjacency matrix $\bm{Y}'$, we exchange the initially estimated parameters with anomalous either in all flows or only in flows being part of $c_1$.

\begin{figure}
    \centering
    \mbox{} 	\small  Change in $\alpha + 0.2$ (Original $\alpha = 0.2$) \mbox{}\\
    \vspace{-0.05cm}
			\includegraphics[scale=0.5, trim= 0 0 0 1.28cm,clip]{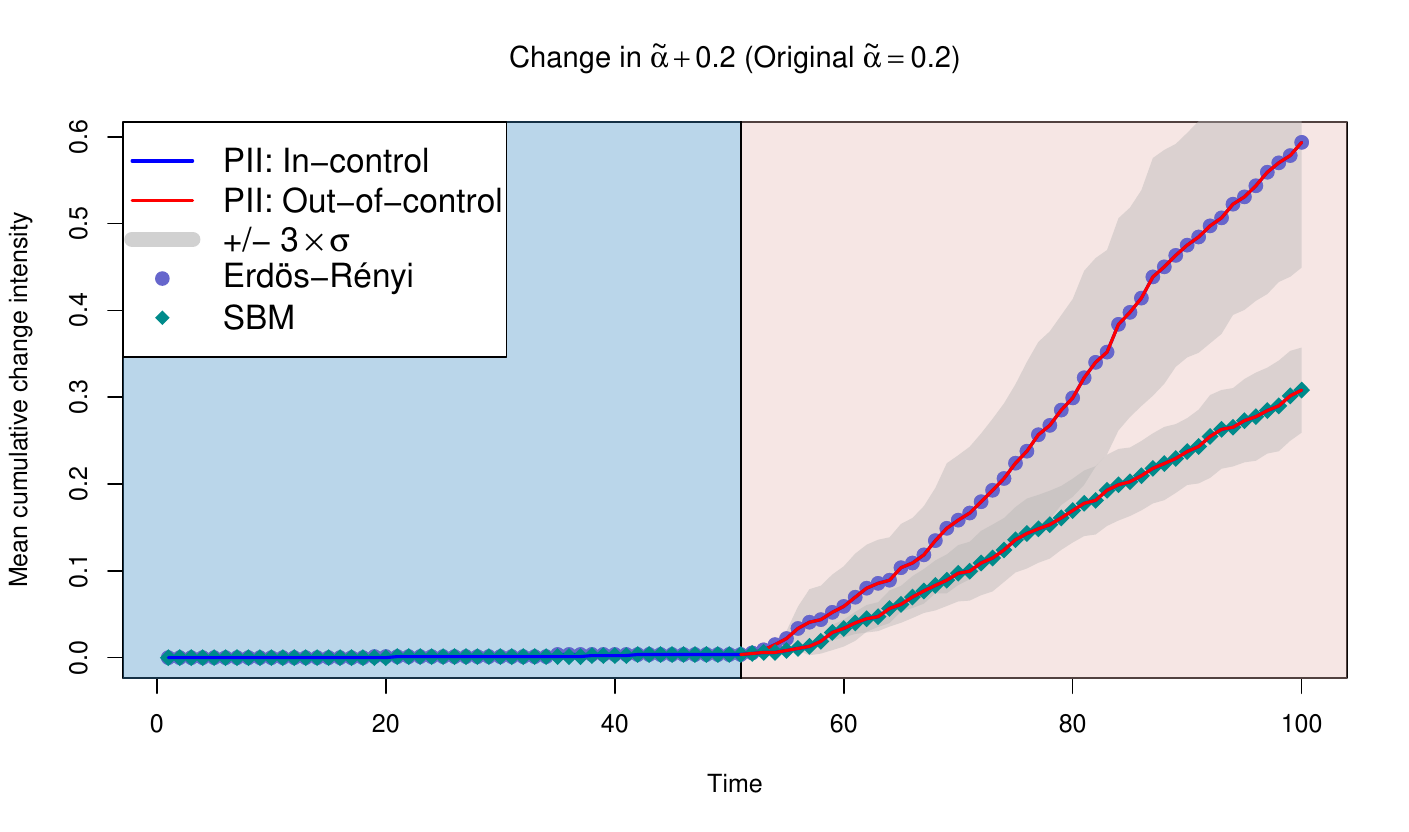}\\
			 \mbox{} 	\small  Change in $\alpha + 0.3$ (Original $\alpha = 0.2$) \mbox{}\\
    \vspace{-0.05cm}
				\includegraphics[scale=0.5, trim= 0 0 0 1.28cm,clip]{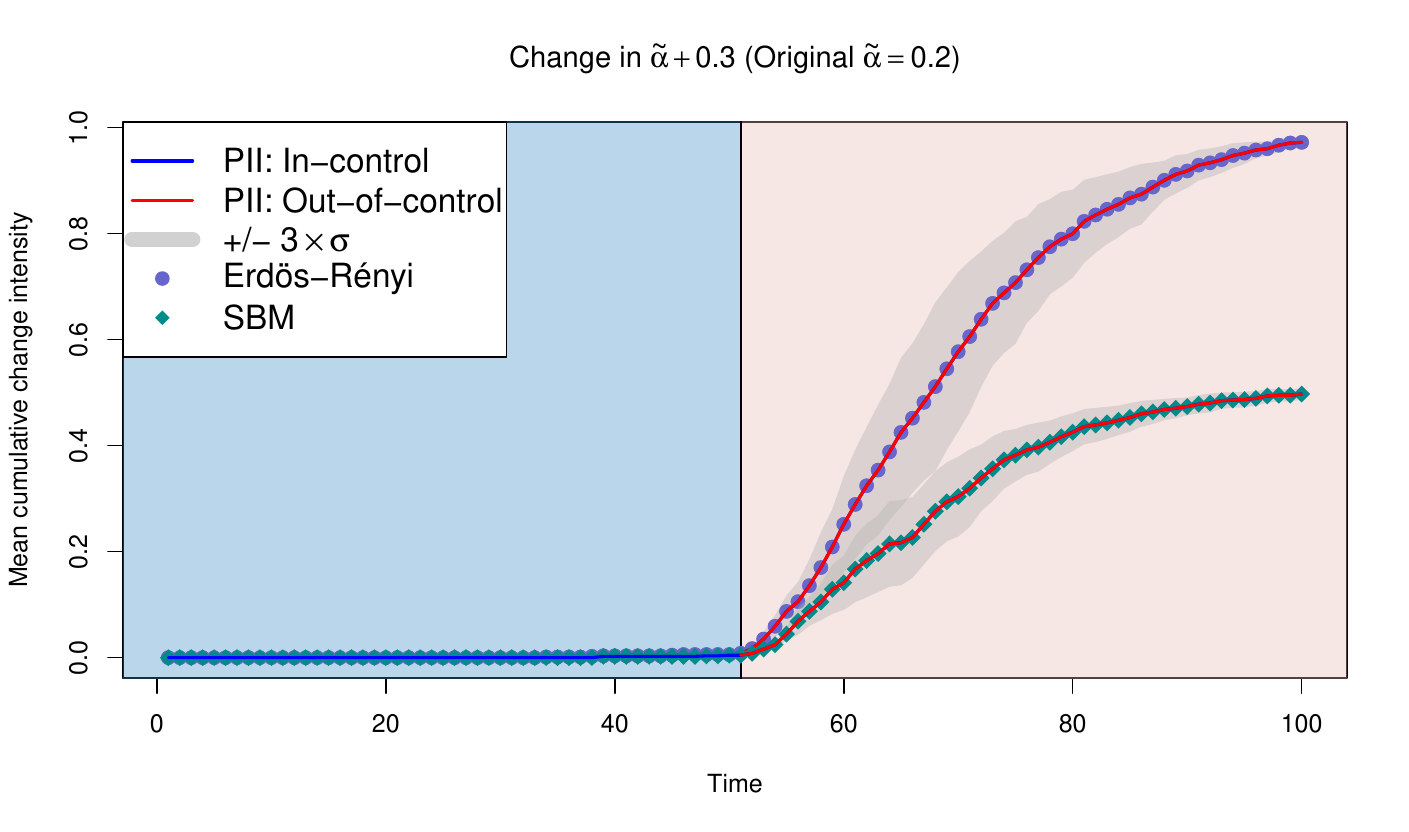}\\
		\mbox{} 	\small  Change in $\alpha + 0.5$ (Original $\alpha = 0.2$) \mbox{}\\		 \vspace{-0.05cm}
					\includegraphics[scale=0.5, trim= 0 0 0 1.28cm,clip]{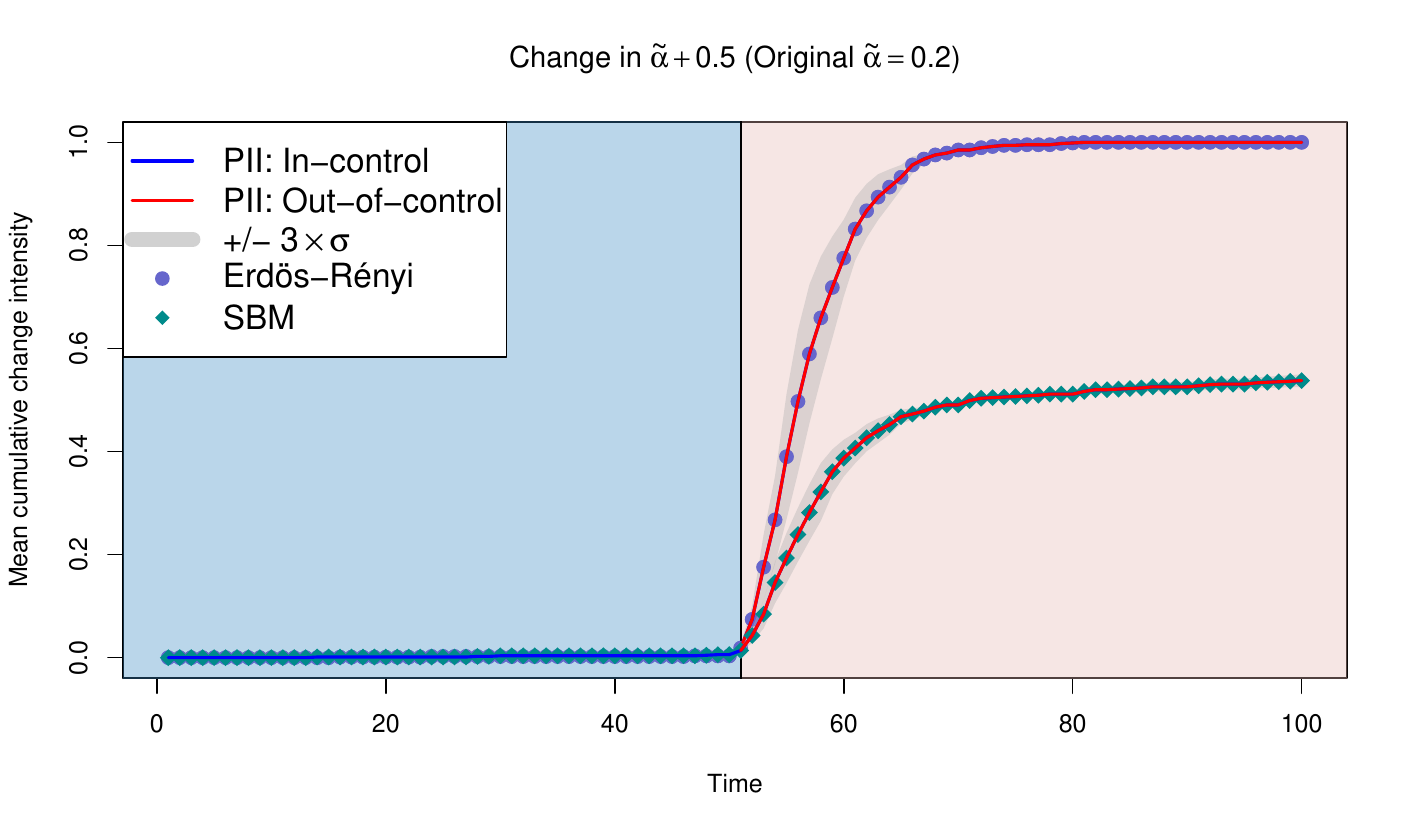}
	 \caption{\centering Simulation study: Visualisation of the mean of cumulative change intensity functions $I_{\bm{X}} (T)$ over 500 iterations with differences in $\alpha$.}
    \label{fig:TENalpha}
\end{figure}

\begin{figure}
    \centering
     \mbox{} 	\small  Change in $\beta + 0.3$ (Original $\beta = 0.3$) \mbox{}\\
    \vspace{-0.05cm}
			\includegraphics[scale=0.5, trim= 0 0 0 1.35cm,clip]{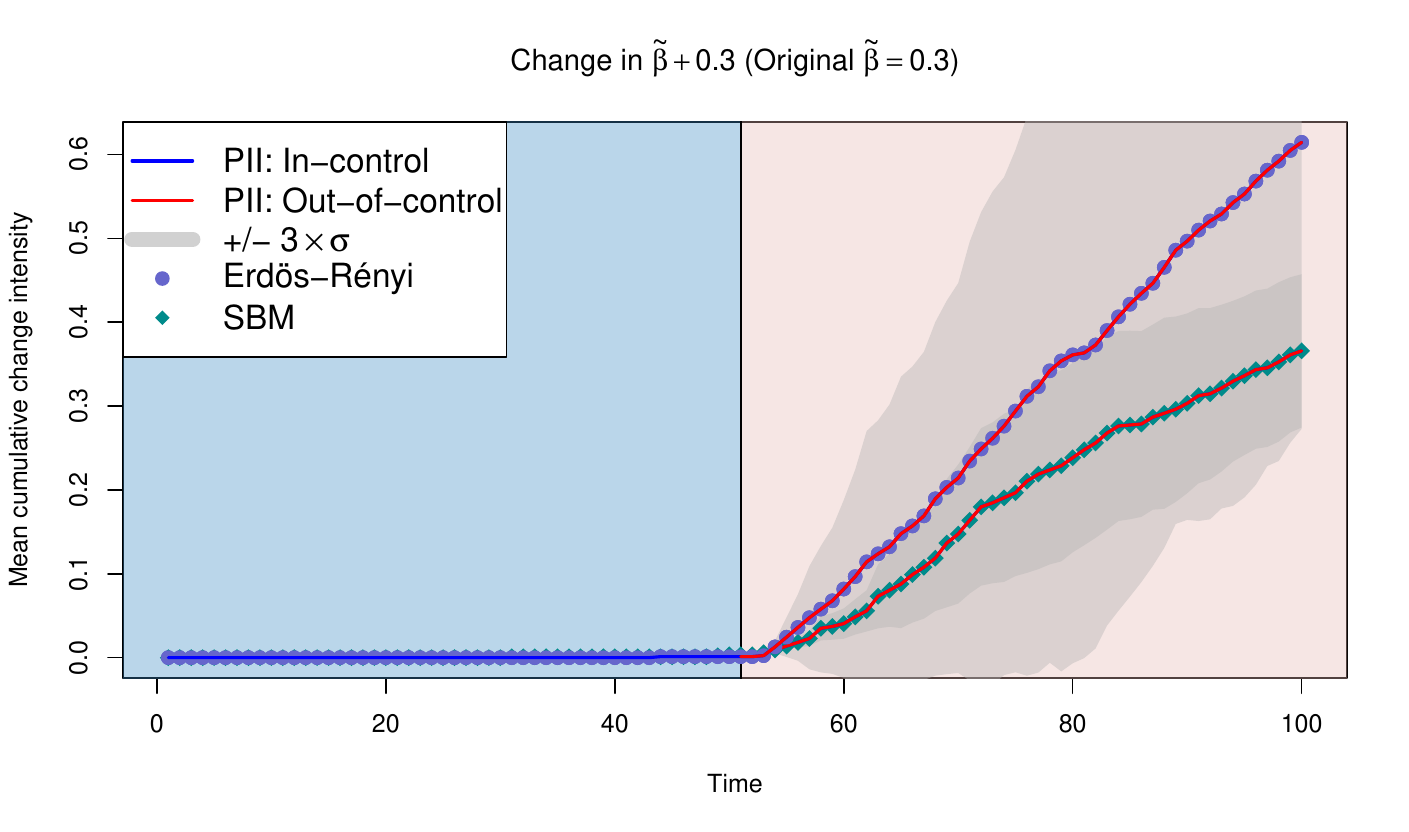}\\
			 \mbox{} 	\small  Change in $\beta + 0.4$ (Original $\beta = 0.3$) \mbox{}\\
    \vspace{-0.05cm}
				\includegraphics[scale=0.5, trim= 0 0 0 1.35cm,clip]{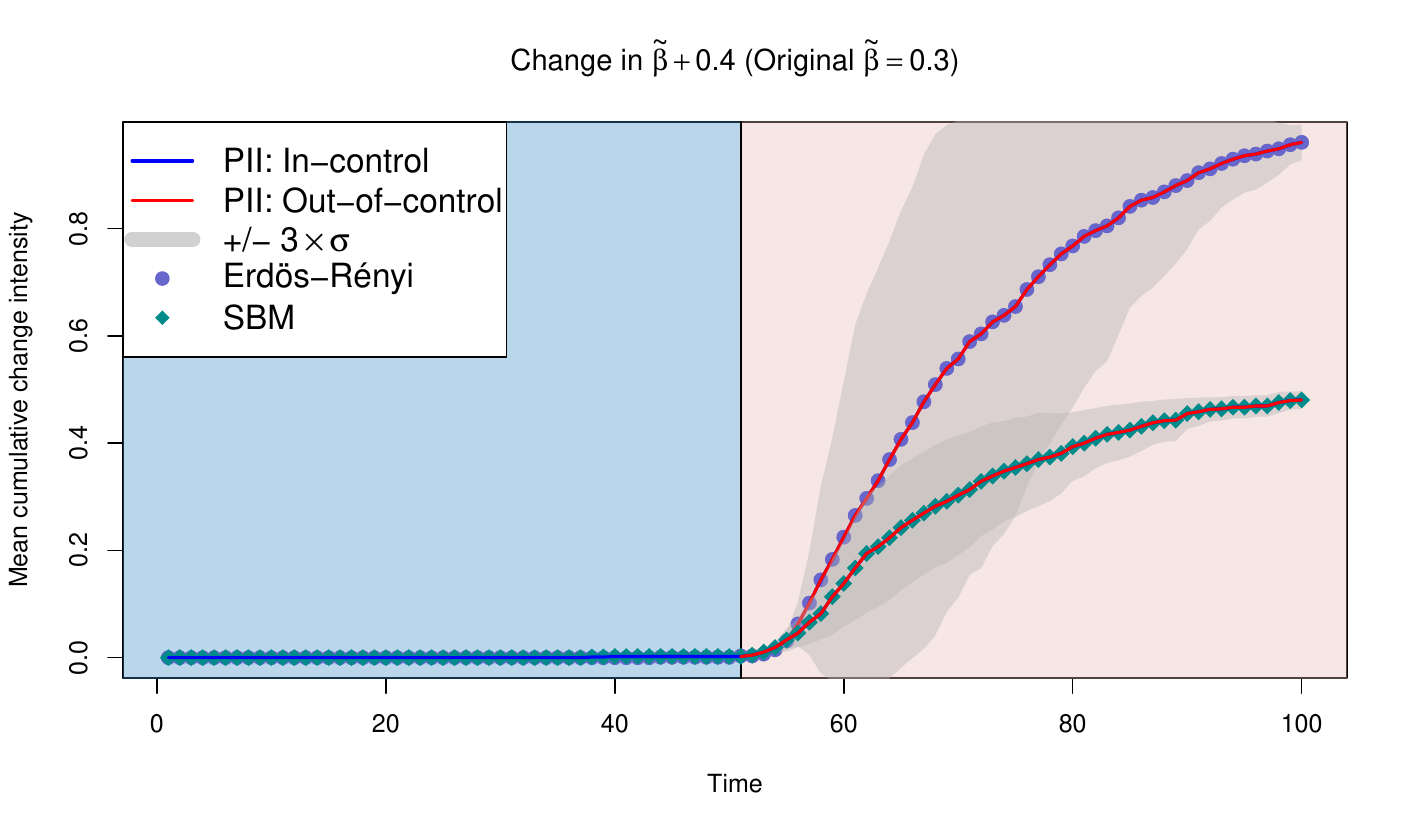}\\
					 \mbox{} 	\small  Change in $\beta + 0.5$ (Original $\beta = 0.3$) \mbox{}\\
    \vspace{-0.05cm}
					\includegraphics[scale=0.5, trim= 0 0 0 1.35cm,clip]{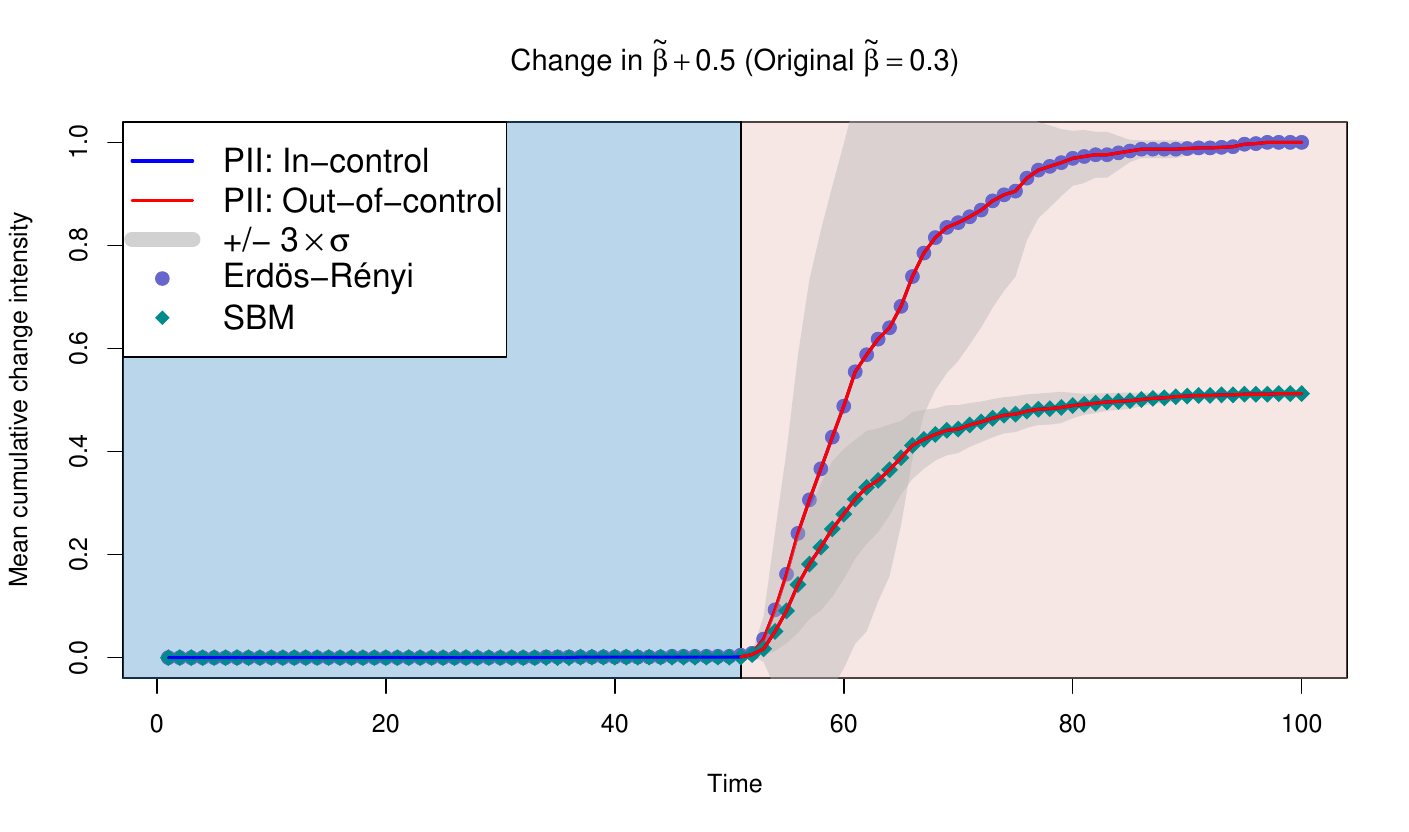}
	 \caption{\centering Simulation study: Visualisation of the mean  of cumulative change intensity functions $I_{\bm{X}} (T)$ over 500 iterations with differences in $\beta$. }
    \label{fig:TENbeta}
\end{figure}

\begin{figure}
    \centering
        \mbox{} 	\small  Change in $\gamma_1 + 1.0$ (Original $\gamma_1 = 2.0$) \mbox{}\\
    \vspace{-0.05cm}
			\includegraphics[scale=0.5, trim= 0 0 0 1.35cm,clip]{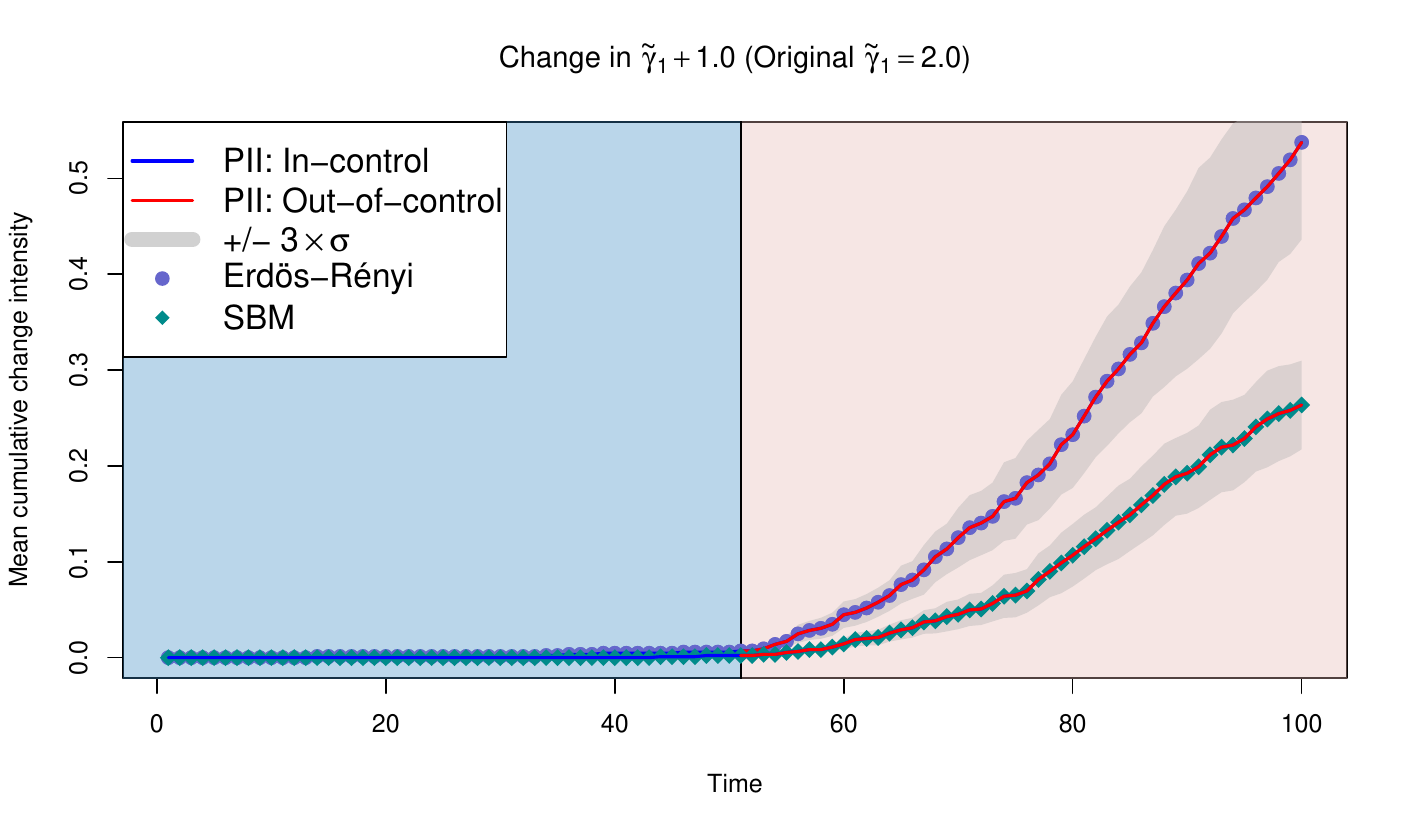}\\
			  \mbox{} 	\small  Change in $\gamma_1 + 1.5$ (Original $\gamma_1 = 2.0$) \mbox{}\\
    \vspace{-0.05cm}
				\includegraphics[scale=0.5, trim= 0 0 0  1.35cm,clip]{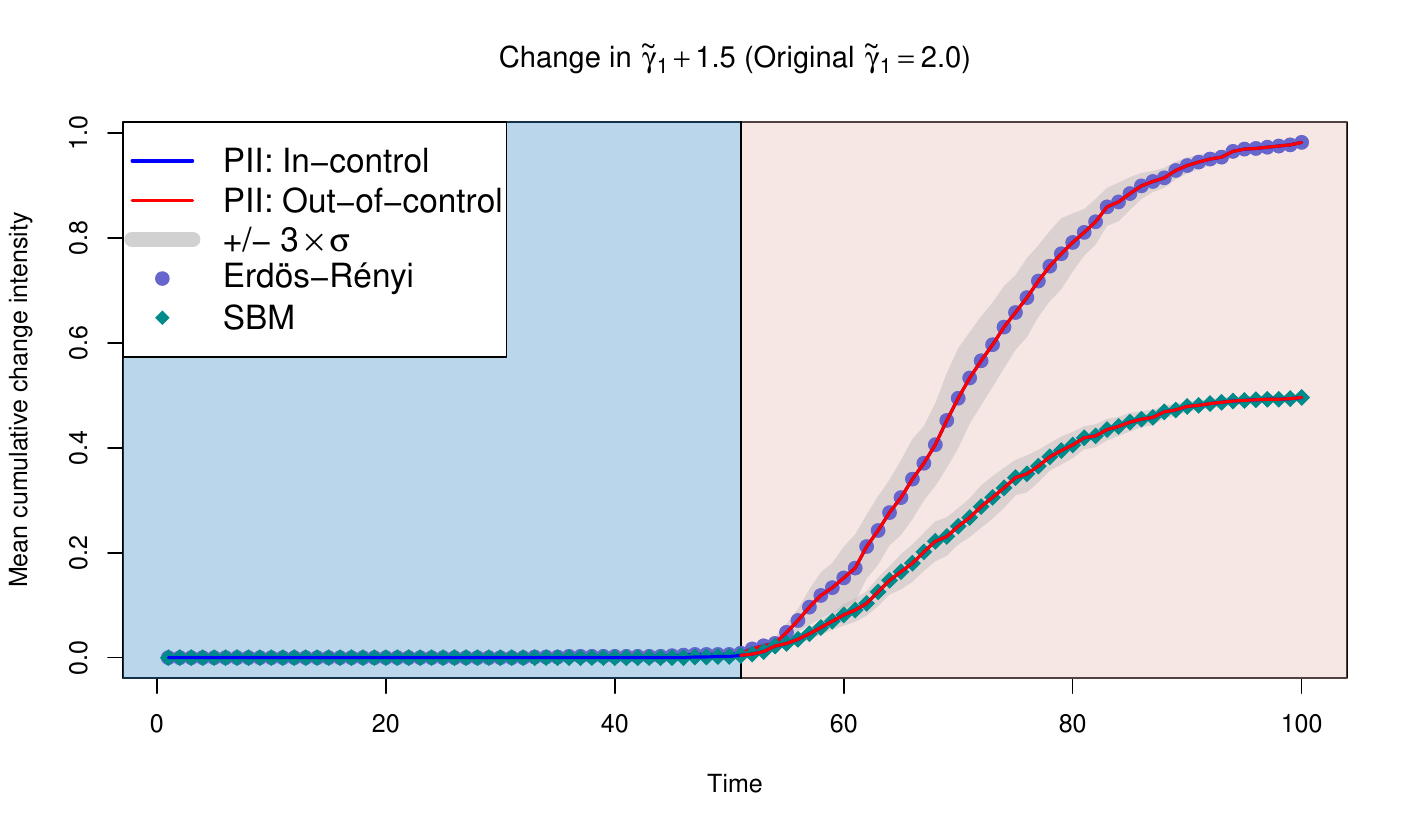}\\
				 \mbox{} 	\small  Change in $\gamma_1 + 2.0$ (Original $\gamma_1 = 2.0$) \mbox{}\\
    \vspace{-0.05cm}
					\includegraphics[scale=0.5, trim= 0 0 0 1.35cm,clip]{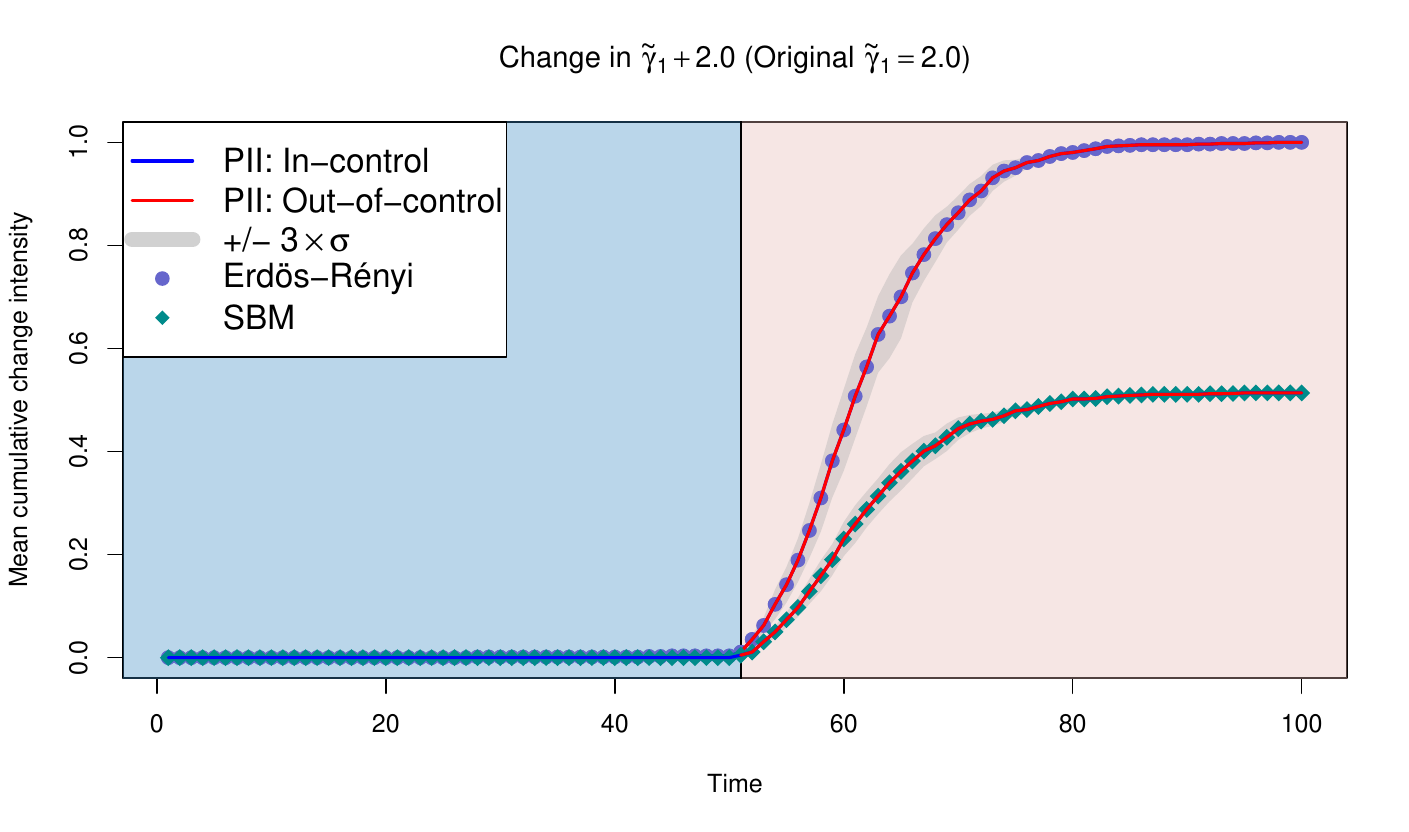}
	 \caption{\centering Simulation study: Visualisation  of the mean of cumulative change intensity functions $I_{\bm{X}} (T)$ over 500 iterations with differences in $\gamma_1$. }
    \label{fig:TENgamma}
\end{figure}

In the following Figures \ref{fig:TENalpha}, \ref{fig:TENbeta} and \ref{fig:TENgamma} we display the monitoring results in Phase II (PII). The blue area defines the in-control and the red the out-of-control part. First, in all plots, we notice the same pattern: The curve that shows the simulation with $\bm{Y}'$ sampled from the SBM stays below another curve that corresponds to the setting with an adjacency matrix generated from the Erd\H{o}s-R\'enyi model. That corresponds to the way the change in the effects was implemented: While in the case of the SBM structure, only one community (half of the flows) was affected by anomalous parameter sizes, the simulation that involves the Erd\H{o}s-R\'enyi structure experiences the change in the complete network, i.e. all flows were affected. Thus, the cumulative intensity change reaches a value of only 0.5 which is the normalised size of a community with 5 flows that presents anomalous behaviour when using the SBM. Second, we clearly recognise that the most difficult type of anomaly for the proposed approach is the change in the neighbourhood effects captured by the parameter $\beta$. It can be seen from the wider uncertainty span and longer run length to detect the insertion of a change in $\beta$. Overall, the simulation study reveals that the monitoring approach is highly effective, meaning that no signals occur when no actual change has been introduced but is only efficient in detecting quickly the time stamp after the change if we are aware of the underlying structure of $G'$ or aim to detect medium or large process innovations.    

\begin{figure}
    \centering
    	 \mbox{} 	\small  Change in $\alpha + 0.3$ (Original $\alpha = 0.2$) \mbox{}\\
    \vspace{-0.05cm}
			\includegraphics[scale=0.45, trim= 0 0 0 1.28cm,clip]{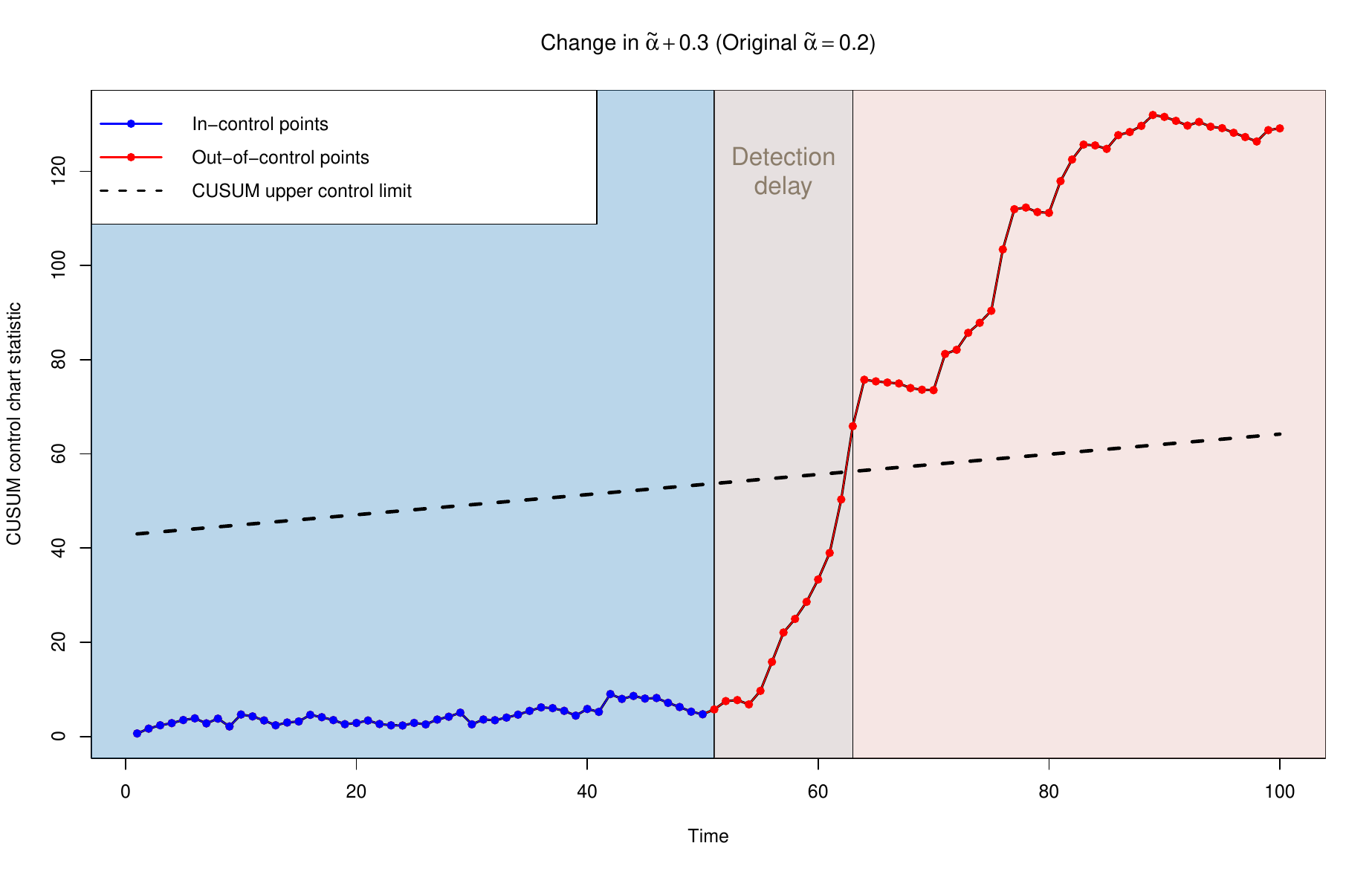}
			
			\caption{\centering Simulation study: Example of monitoring one specific flow from a TEN process.}
  \label{fig:CUSUM}
\end{figure}

In case one specific flow reflects the state of the whole system or is particularly relevant for the flawless functionality of a network, we could focus solely on its monitoring. For that, we would apply the CUSUM control chart and directly determine the change point. Visually, a possible outcome could look as displayed in Figure \ref{fig:CUSUM}.

In the following section, as a real-world illustration, we present the monitoring of cross-border physical electricity flows across Europe.

\section{Monitoring of Cross-Border Physical Electricity Flows}
\label{sec:ele}
Cross-country analysis of electricity trade, especially in the context of renewable energy, is a relevant field of study that i.a. contributes to the evaluation of transmission infrastructure policies (cf. \citealp{abrell2016cross}). It is worth noting that the cross-border physical flow network indicates the actual flow of electricity that corresponds to the laws of physics and could differ from the scheduled commercial exchanges which reflect the economic relations between the market parties and, therefore, be of particular technical importance, allowing for the flawless electricity trade across Europe. To the best knowledge of the authors, there are no studies related to the change point detection in European Cross-Border Physical Flows (CBPFs), so we illustrate the monitoring of these processes below. 

\subsection{Data Description}

European Network of Transmission System Operators for Electricity (ENTSO-E) provides data about the cross-border physical flow and scheduled commercial electricity exchanges of more than 40 countries\footnote{Central collection and publication of electricity generation, transportation and consumption data and information
for the pan-European market. ENTSO-E Transparency platform (2023).  \url{https://transparency.entsoe.eu}}. It offers the data for each hour for years starting from 2014. However, after careful investigation of the amount of missing data, we decided to concentrate on the period from January 1, 2018, to November 27, 2022, and use a weekly aggregation of the electricity values given in megawatts.

\begin{figure}[!tbp]
  \centering
  \vspace{-1.3cm}
  \begin{minipage}[b]{0.9\textwidth}
  \hspace{2cm}
    \includegraphics[scale=0.5, trim= 5cm 0 5cm 0,clip]{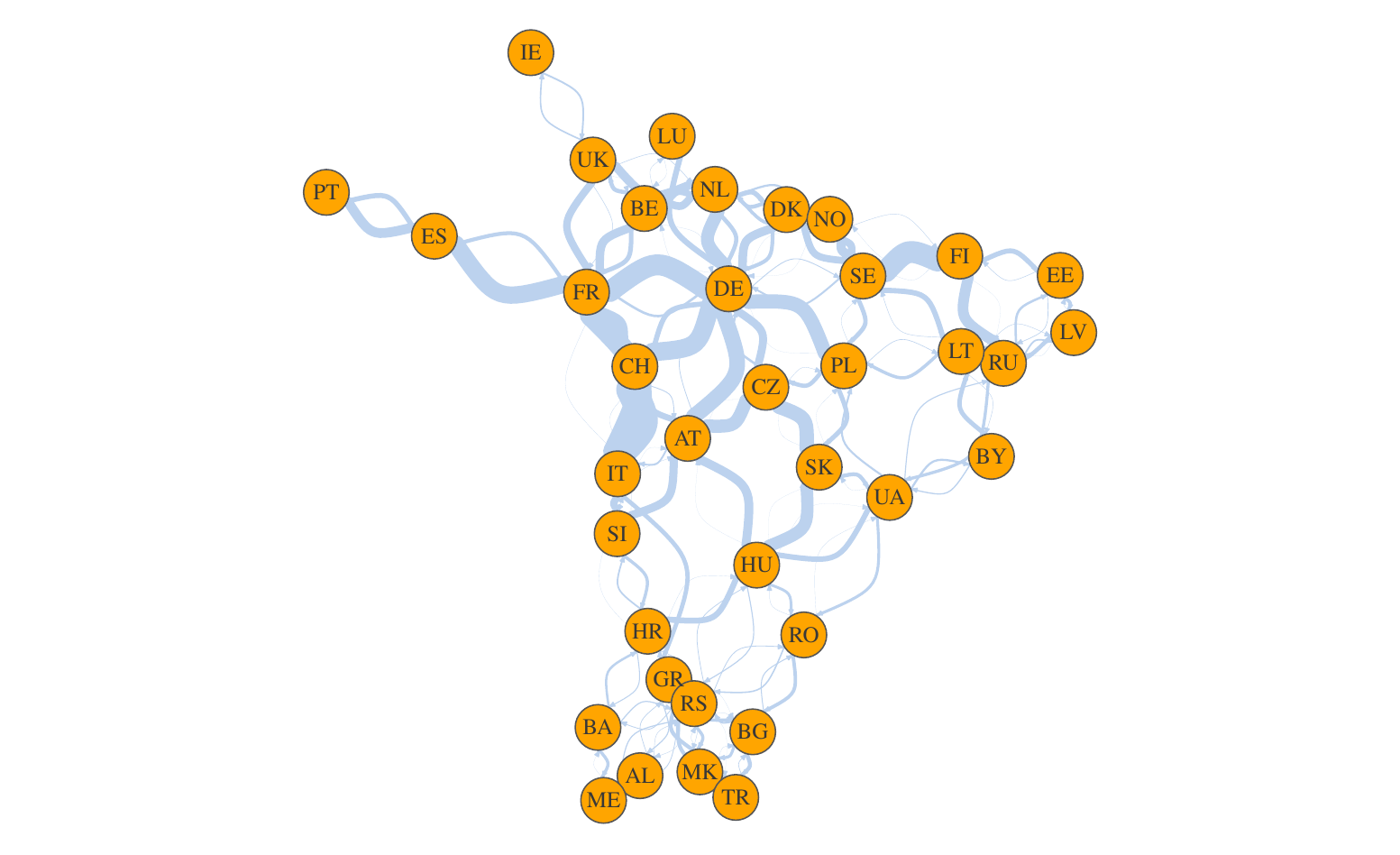}\\
    \centering  (a) The edge thickness implicates the strength of the flow.
  \end{minipage}
 \vfill
  \begin{minipage}[b]{0.9\textwidth}
  \hspace{0.8cm}
    \includegraphics[scale=0.45, trim= 2cm 4cm 2cm 3cm,clip]{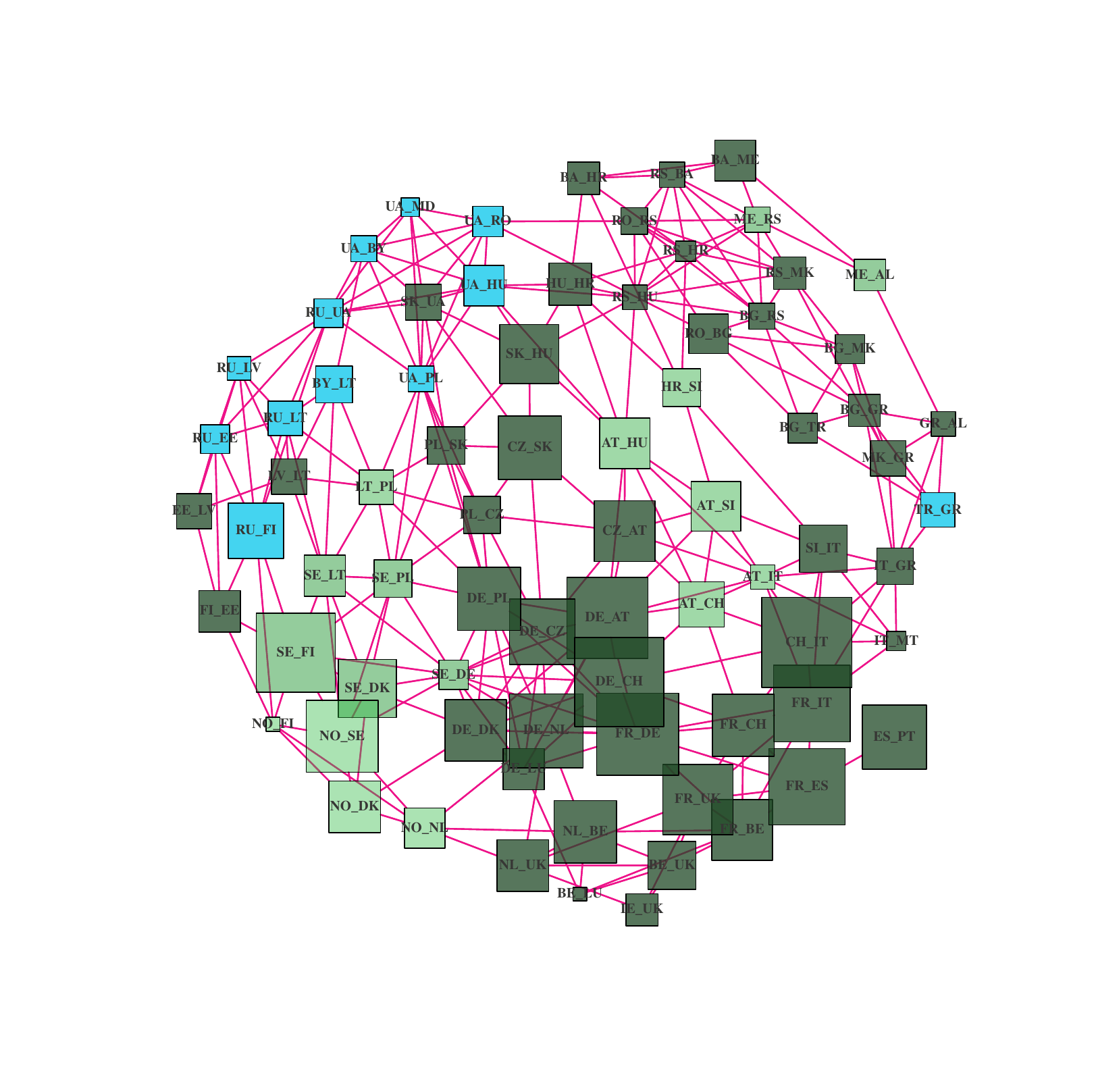}\\
    \centering (b) The size of the nodes implicates the strength of the electricity exchange and the colour its proportion of electricity generated with renewable energy sources (light green corresponds to the higher proportion, blue corresponds to the missing information).
  \end{minipage}
  \caption{\centering Conventional representation of the cumulated cross-border physical electricity flow in the year 2019 (a) and new representation (b) following the description in Section \ref{subsubsec:GNARXTEN}.}
  \label{fig:CBPF}
\end{figure}

The graph in Figure \ref{fig:CBPF} (a) represents the aggregated quantity of electricity
which was transmitted in the year 2019 so that the wideness of edges reflects the
total amount of power which was passed between the two countries. The higher this value
is, the wider the edge between two nodes. However, what we can also notice is the existence of parallel edges, i.e. one flow $f_1$ coming from France (FR) to Spain (ES) and another flow  $f_2$ back. If we decide to directly apply a new representation as introduced in Section \ref{subsubsec:GNARXTEN}, we would obtain a network that would be considerably bigger than the original graph. Hence, we need to aggregate both flows $f_1$ and $f_2$ and take one vertex to represent a country pair. There are different possibilities for doing it, we decide on three of them and discuss each of them subsequently.

\subsection{Phase I Modelling}

To avoid a considerable expansion in the new representation of the CBPFs (see Section \ref{subsubsec:GNARXTEN}), we introduce three different aggregation strategies of both flows $f_1$ and $f_2$, fitting and comparing three models, respectively. The first statistic is the Box-Cox transformation of the sum of both flows $\mathscr{M}_1 = ln(f_1 + f_2 + 1)$ with $\lambda_1 = 0$ and $\lambda_2 = 1$ (cf. \citealp{box1964analysis}) that reflects the overall strength of the exchange. The second statistic measures the asymmetry in both flows as the difference of the Box-Cox transformed variables $f_1$ and $f_2$, i.e. $\mathscr{M}_2 = \ln(f_1 + 1) - \ln(f_2 + 1)$. It implies whether it is more common for one country to import or export electricity. The third statistic captures proportionally differences in both flows, being $\mathscr{M}_3 = \frac{f_1 - f_2}{f_1 + f_2}$. No additional transformation is required in this case as $\mathscr{M}_3 \in [-1, 1]$\textbackslash$\{0\}$. Figure \ref{fig:CBPF}  (b) illustrates a new representation of CBPFs, where the new adjacency structure is constructed using the rules of a line graph described in Section \ref{subsubsec:GNARXTEN}.

As no exceptional events are known in year 2018 and 2019 that could considerably disturb the CBPF network, we decide to choose these two years as Phase I. Testing different model settings, the final set-up involves a global autoregressive parameter $\Tilde{\alpha}$ with $p = 1$, 1-stage neighbourhood $\Tilde{\beta}$ and two covariates that correspond to the Box-Cox transformed total amount of electricity generated by renewable energy sources from both countries with the effects $\Tilde{\gamma_1}$ and $\Tilde{\gamma_2}$, having $\bm{q} = (0, 0)$.

Table \ref{table:coefficients} presents the results of estimating the parameters for each choice of the aggregation statistic $\mathscr{M}$ using 7828 observations ($76\times (52\times 2 - p)$), where 76 defines the number of bilateral exchanges. As we can observe, the GNARX model fits the data in Phase I well and confirms the relevance of accounting for the network structure as well as external effects when modelling TEN processes. Using the displayed coefficients, we proceed with the monitoring of Phase II which consists of the years 2020--2022. 

\begin{table}
\begin{center}
\begin{tabular}{l c c c }
\hline
 & Model $\mathscr{M}_1$ & Model $\mathscr{M}_2$  & Model $\mathscr{M}_3$  \\
\hline
$\Tilde{\alpha}$     & \bm{$0.896^{***}$} & \bm{$0.904^{***}$}  & \bm{$0.910^{***}$}  \\
           & $(0.005)$     & $(0.005)$      & $(0.005)$      \\
$\Tilde{\beta}$    & \bm{$0.094^{***}$} & \bm{$0.019^{\cdot}$}        & \bm{$0.020^{\cdot}$}        \\
           & $(0.006)$     & $(0.011)$      & $(0.011)$      \\
$\Tilde{\gamma}_1$  & \bm{$0.003^{\cdot}$}       & \bm{$-0.034^{***}$} & \bm{$-0.003^{***}$} \\
           & $(0.002)$     & $(0.006)$      & $(0.001)$      \\
$\Tilde{\gamma}_2$  & \bm{$0.005^{\cdot}$}      & \bm{$0.032^{***}$}  & \bm{$0.003^{***}$}  \\
           & $(0.002)$     & $(0.006)$      & $(0.001)$      \\
\hline
$\Bar{\mathscr{M}}$        & 10.84         & 0.14          & 0.03          \\
$\Tilde{\sigma}_\mathscr{M}$          & 2.00         & 6.81          & 0.84         \\
RMSE       & 0.89         & 2.75          & 0.34          \\
\hline
\multicolumn{4}{l}{\scriptsize{$^{***}p<0.001$, $^{**}p<0.01$, $^*p<0.05$,  $^{\cdot}p\leq0.1$}}
\end{tabular}
\caption{\centering Phase I modelling.}
\label{table:coefficients}
\end{center}
\end{table}

\subsection{Phase II Monitoring}

\begin{figure}
    \centering
			\includegraphics[scale=0.35, trim= 0 0 0 0,clip]{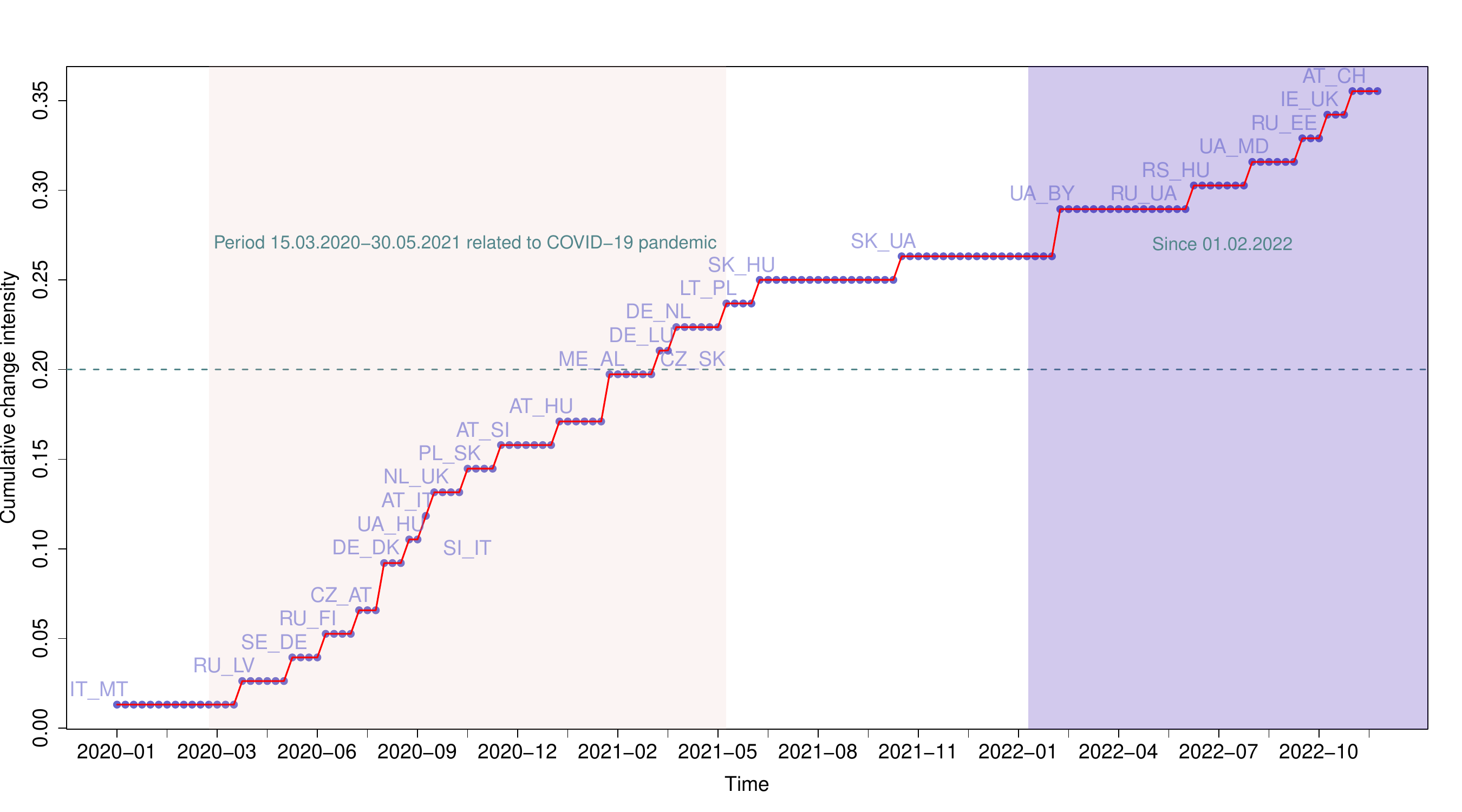}
			\caption{\centering Results of monitoring $\mathscr{M}_1$, obtaining in total 27 change points.}
  \label{fig:M_1}
\end{figure}

\begin{figure}
    \centering
			\includegraphics[scale=0.35, trim= 0 0 0 0,clip]{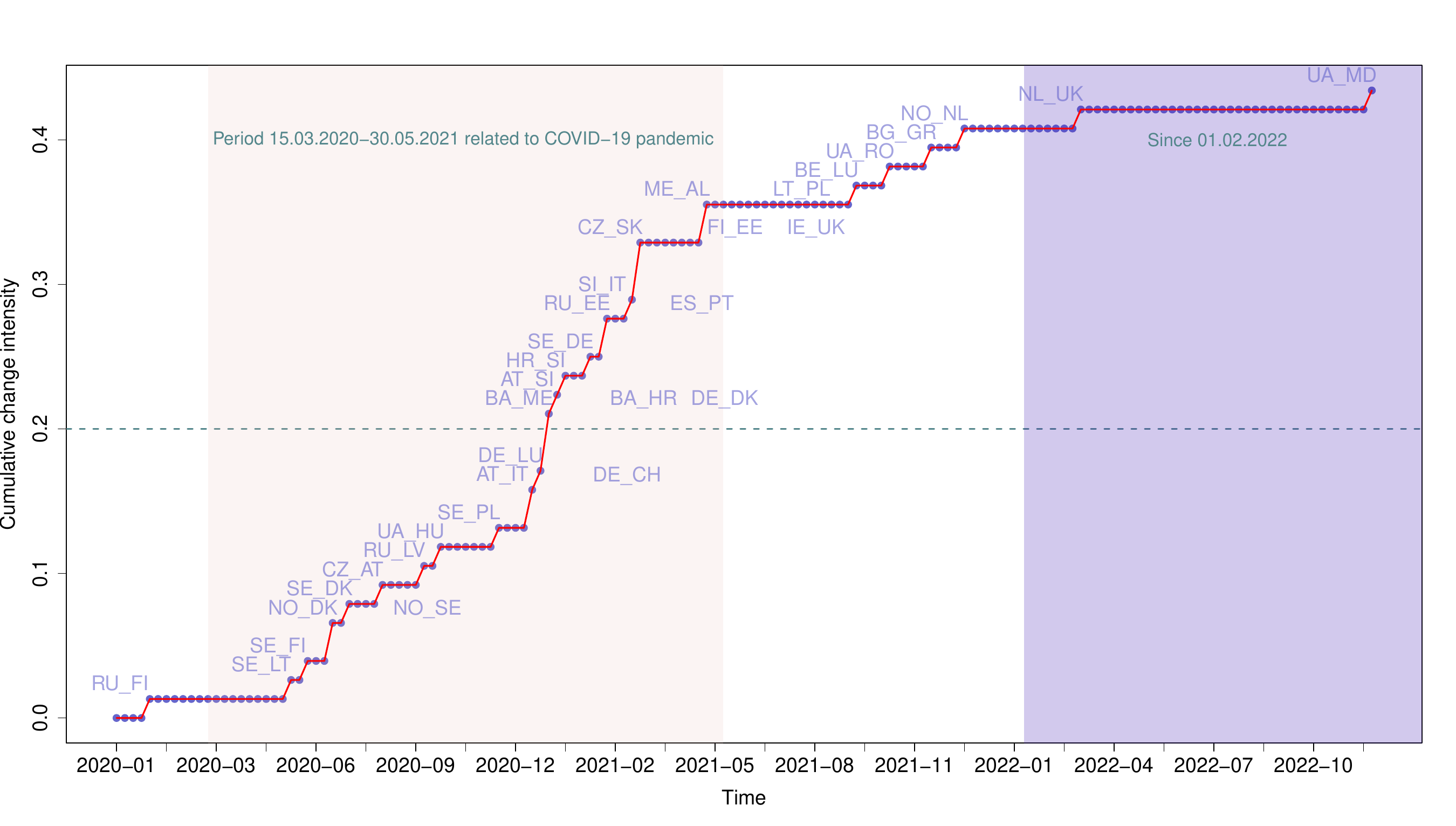}
			\caption{\centering Results of monitoring $\mathscr{M}_2$, obtaining in total 33 change points.}
  \label{fig:M_2}
\end{figure}

\begin{figure}
    \centering
			\includegraphics[scale=0.35, trim= 0 0 0 0,clip]{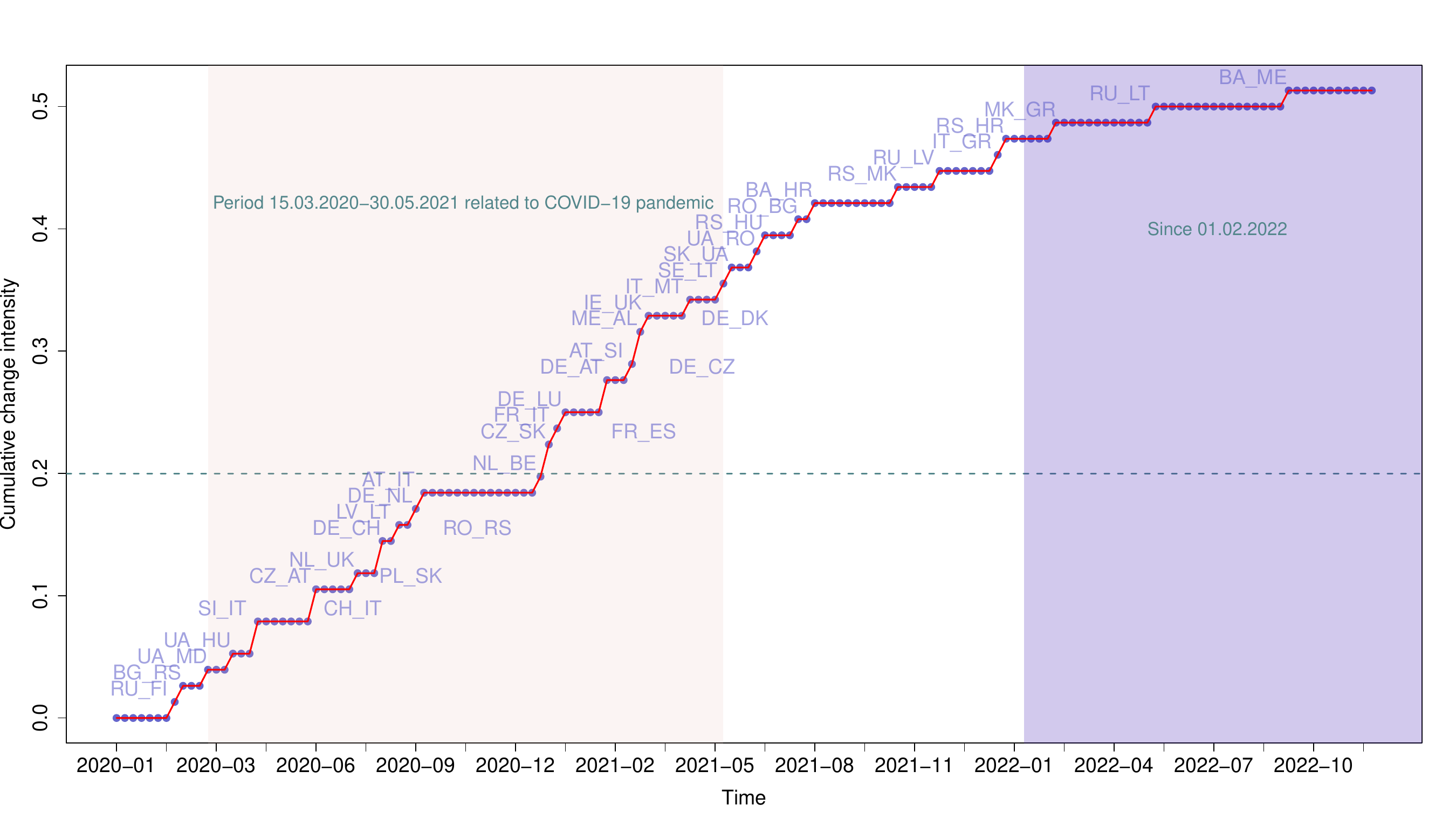}
			\caption{\centering Results of monitoring $\mathscr{M}_3$, obtaining in total 39 change points.}
  \label{fig:M_3}
\end{figure}

Figures \ref{fig:M_1}, \ref{fig:M_2} and \ref{fig:M_3} display the monitoring results during Phase II. Out of 76 country pairs, between 27 and 39 pairs have had a change point according to the implemented CUSUM control charts based on either $\mathscr{M}_1$, $\mathscr{M}_2$ or $\mathscr{M}_3$ statistics. We highlight two periods during Phase II: The period related to severe situations due to the COVID-19 pandemic and the period starting in February 2022 where several crises are expected or happened due to the Russian-Ukrainian war. Considering the threshold $W$, we define it to be $W = 0.2$, i.e. when in at least 16 bilateral exchanges a change point is detected.

Comparing three different aggregation methods, we notice similarities in detecting changes during the pandemic. However, starting from June 2021 the behaviour of the control charts significantly differs. In the case of the results with $\mathscr{M}_1$, the time between two defined periods remains relatively stable compared to two other charts with $\mathscr{M}_2$ and $\mathscr{M}_3$. Then, the first detection on March 6, 2022 (aggregating the days February 28 -- March 6) considers the beginning of the war between Russia and Ukraine. Surprising could be the fact that another country pair experiences a change in the same week, namely Belarus and Ukraine. Later, another change detection occurs at the end of May in the pair Russia and Estonia,  reflecting the political agreement of the Baltic states to stop any energy trade with Russia. The same date and the reason are also given in the control chart with $\mathscr{M}_3$ in the pair Russia and Lithuania\footnote{https://enmin.lrv.lt/en/news/no-more-russian-oil-gas-and-electricity-imports-in-lithuania-from-sunday}.  In terms of the signal in the pair Ukraine and Moldova (control chart with $\mathscr{M}_2$), it could be related to the launch of the planned commercial exchanges announced at the end of June 2022 and further being increased in the beginning of autumn \footnote{https://www.entsoe.eu/news/2022/09/04/transmission-system-operators-of-continental-europe-decide-to-further-increase-trade-capacity-with-the-ukraine-moldova-power-system/}.

Overall, we can notice that without expert knowledge of the CBPF network, it is challenging to reflect a particular reason for the change point, however, we also can see a reliable performance of the proposed framework in detecting major events such as the pandemic and the energy crisis related to the begin of the Russian-Ukrainian war. The caveat to be aware of is the interpretation of the chosen aggregation statistic $\mathscr{M}_1, \mathscr{M}_2, \mathscr{M}_3$ as in some of the cases the signalled anomalies considerably differ.

\section{Conclusion}
The network with a given structure but a random process on its edges that we define in this work as a Temporal Edge Network (TEN) can be of particular interest for guaranteeing the safety of the infrastructure but also for foreseeing possible accidents. In this manuscript, we present the monitoring framework to detect anomalies in TENs by combining the GNARX model and the CUSUM control chart based on residuals.

There might still exist the question of why monitoring networks with a fixed structure needs special treatment. For example, why not try to ``randomise'' the connections? The explanation is straightforward: It could be possible to select a flow threshold for creating dynamics in the graph by deleting or adding the links, however, the anomalies which would be detected in this case are similar to those which are the focus of the random network monitoring. In our case, we are rather interested in anomalies occurring in the edge process, e.g. changes in the flow's strength or some other temporal deviations. Hence, we introduce a new way of representing TEN processes where edges become the main focus of both modelling and monitoring parts.

The proposed change point detection framework is applied only to the TENs observed at discrete times. It is an open research field on how and when to extend the monitoring to continuous times. Equally important is to research when different adjacency matrices for different time points should be introduced and whether it will benefit the overall monitoring procedure. Also, for monitoring a TEN process at once, we have introduced a cumulative change intensity function. However, an appealing way to perform such monitoring would be a suitable multivariate control chart or another statistical monitoring tool that allows for medium-sized multivariate processes.

As an empirical illustration, we have monitored a network of bilateral electricity flows across Europe. Beyond that, our framework could potentially be applied for monitoring traffic flows on roads or rail systems to optimise transportation infrastructure as well as within the computer communication domain, aiding in cybersecurity. In an environmental context, TENs may be useful in evaluating water transfers between areas or companies, contributing to sustainable water resource management. 

This work is devoted to networks with static structures that fit well with the considered process on edges being cross-border physical electricity flow. However, thinking of potential processes with a changing underlying structure over time, it is important to determine whether the monitoring framework would experience substantial changes in terms of re-estimating the parameters in Phase I or challenges in performing multivariate monitoring when some of nodes or edges disappear with time. Moreover, the effectiveness of the proposed monitoring procedure strongly relies on the assumption of the GNARX model being suitable to the considered data. Thus, in case the data cannot be well represented by the GNARX model, i.e. the count data are given, an alternative modelling approach or an extension to the currently existing model is required.

\end{document}